\documentclass[11pt,preprint]{aastex} 

\usepackage{natbib,amsmath}
\usepackage{color}

\definecolor{orange}{cmyk}{0,0.4,0.8,0.2}
\definecolor{darkorange}{rgb}{.71,0.21,0.01}
\definecolor{darkgreen}{rgb}{.12,.54,.11}

\usepackage{hyperref}
\hypersetup{pdftex,  
  breaklinks=true,  
  colorlinks=true,
  urlcolor=blue,
  linkcolor=darkorange,
  citecolor=darkgreen,
}

\def\lsim{\hbox{ \rlap{\raise 0.425ex\hbox{$<$}}\lower 0.65ex\hbox{$\sim$}}}
\def\gsim{\hbox{ \rlap{\raise 0.425ex\hbox{$>$}}\lower 0.65ex\hbox{$\sim$}}}
\def\arcmin{\hbox{$^\prime$}}
\def\arcsec{\hbox{$^{\prime\prime}$}}
\def\arcdeg{\mbox{$^\circ$}}%

\def\fh{\hbox{$~\!\!^{\rm h}$}}
\def\fm{\hbox{$~\!\!^{\rm m}$}}
\def\fs{\hbox{$~\!\!^{\rm s}$}}
\def\ale{\mathrel{\hbox{\rlap{\hbox{\lower4pt\hbox{$\sim$}}}\hbox{$<$}}}}
\def\age{\mathrel{\hbox{\rlap{\hbox{\lower4pt\hbox{$\sim$}}}\hbox{$>$}}}}

\begin{document}
\title{Automating Discovery and Classification of Transients and Variable Stars in the Synoptic Survey Era}

\def\berk{1}
\def\stat{2}
\def\lbnl{3}
\def\caltech{4}
\def\weiz{5}
\def\dun{6}

\author{J. S. Bloom\altaffilmark{\berk}, J. W. Richards\altaffilmark{\berk,\stat},
        P. E. Nugent\altaffilmark{\lbnl,\berk}, R. M. Quimby\altaffilmark{\caltech}, 
        M. M. Kasliwal\altaffilmark{\caltech}, 
        D.~L. Starr\altaffilmark{\berk}, D. Poznanski\altaffilmark{\berk,\lbnl}, 
        E. O. Ofek\altaffilmark{\caltech}, S. B. Cenko\altaffilmark{\berk}, 
        N.~R. Butler\altaffilmark{\berk}, S. R. Kulkarni\altaffilmark{\caltech}, A.  Gal-Yam\altaffilmark{\weiz}, N. Law\altaffilmark{\dun}}

\affil{$^\berk$ Department of Astronomy, 
        University of California, Berkeley, CA 94720-3411.}

\affil{$^\stat$ Department of Statistics, 
        University of California, Berkeley, CA 94720-7450.}

\affil{$^\lbnl$ Computational Cosmology Center, Lawrence Berkeley National Laboratory, 1 Cyclotron Road, Berkeley, CA 94720, USA}

\affil{$^\caltech$ Cahill Center for Astrophysics, California Institute of Technology, Pasadena, CA, 91125, USA}

\affil{$^\weiz$ Department of Particle Physics and Astrophysics, Faculty
of Physics, The Weizmann Institute of Science, Rehovot 76100,
Israel}

\affil{$^\dun$ Dunlap Institute for Astronomy and Astrophysics, University of Toronto, 50 St. George Street, Toronto M5S 3H4,
Ontario, Canada}

\begin{abstract}
The rate of image acquisition in modern synoptic imaging surveys has already begun to outpace the feasibility of keeping astronomers in the real-time discovery and classification loop. Here we present the inner workings of a framework, based on machine-learning algorithms, that captures expert training and ground-truth knowledge about the variable and transient sky to automate 1) the process of discovery on image differences and, 2) the generation of preliminary science-type classifications of discovered sources. Since follow-up resources for extracting novel science from fast-changing transients are precious, self-calibrating classification probabilities must be couched in terms of efficiencies for discovery and purity of the samples generated. We estimate the purity and efficiency in identifying real sources with a two-epoch image-difference discovery algorithm for the Palomar Transient Factory (PTF) survey.  Once given a source discovery, using machine-learned classification trained on PTF data, we distinguish between transients and variable stars with a 3.8\% overall error rate (with 1.7\% errors for imaging within the Sloan Digital Sky Survey footprint). At $>$96\% classification efficiency, the samples achieve 90\% purity. Initial classifications are shown to rely primarily on context-based features, determined from the data itself and external archival databases. In the $\sim$one year since autonomous operations, this discovery and classification framework has led to several significant science results, from outbursting young stars to subluminous Type IIP supernovae to candidate tidal disruption events. We discuss future directions of this approach, including the possible roles of crowdsourcing and the scalability of machine learning  to future surveys such a the Large Synoptical Survey Telescope (LSST).
\end{abstract}


\section{Introduction}
The arrival of the era of synoptic imaging surveys heralds the start of a new chapter of time-domain astrophysics, where the real-time processing of images taxes the capacity to transport the data from remote sites and pushes to the limit the computational capabilities at processing centers (e.g., \citealt{2011EAS....45..281J}). More profoundly novel, however, is that the data volumes have begun to surpass what is possible to visually inspect by even large teams of astronomers and volunteer ``citizen scientists.'' This necessitates an increasingly more central role of software and hardware frameworks to supplant the traditional roles of humans in the real-time loop.

This abstraction of people away from the logistics of the scientific process has been progressing rapidly, starting  with the acquisition process itself. Indeed, robotic telescopes\footnote{For a list of robotic telescopes currently operating, see \\ \url{http://www.uni-sw.gwdg.de/~hessman/MONET/links.html}.}, capable of taking data autonomously at remote sites, have become an increasingly common form of operation at the sub-meter- and meter-class level (cf.~\citealt{ca2010}). Many robotic systems use queuing algorithms that optimize nightly observing over several scientific programs  and many are capable of being interrupted by external alerts to observe high-priority transients (e.g., \citealt{2001ASPC..246..121F,2002SPIE.4845..126V,2003PASP..115..132A,2006PASP..118.1396C,2006ASPC..351..751B,2008AN....329..321S,2010arXiv1002.0108K}). Data from such facilities can be automatically transported, processed, photometered, and ingested into databases without human intervention. 

Since imaging data has spurious sources of noise and artifacts that can mimic real astrophysical sources, in the absence of watchful trained eyes on the images themselves, autonomous discovery of transients and variable stars on synoptic imaging surveys is a significant challenge. Threshold cuts on photometric quality, changes in apparent magnitudes, etc., are effective in discovering bona fide astrophysics sources \citep{2009ApJ...696..870D,2010AdAst2010E..54S}. However, multi-parameter thresholding tends to be suboptimal because it treats each parameter derived from a given candidate as an independent variable when clearly there can be correlations between parameters. Matched filtering---looking for light curve trends that fit the scientific expectation from a certain class of variables (e.g., microlensing; \citealt{1996AJ....112.2872T,2003MNRAS.341.1373B})---can be a very effective tool to discover new events, but other sorts of variables and transients are not easily recovered from that view of the dataset. Likewise, previous machine-learning based discovery (e.g., supernova discovery with the Supernova Factory; \citealt{2007ApJ...665.1246B}) have been optimized on domain-specific discovery, leaving aside the multitude of other variables not of direct interest to that particular project.  
 
Discovery that a varying source is truly astrophysical does not mean that the origin of that variability is understood. Indeed, while it is tempting to conflate the process of discovery with classification, by making sequential the two decisions, different machineries can be brought to bear on each. The literature on autonomous classification, by various computational techniques, has been growing rapidly; indeed a wide range of machine-learning techniques have been applied to classification of large astronomical datasets (see \citealt{2008AIPC.1082..287M} and \citealt{2011arXiv1104.3142B} for review).  Aside from domain-specific classification (microlensing and supernovae), most work concerns classification of variables stars on historical datasets {\it in retrospect}, where analysis is performed after most of the data have been collected and cleaned (e.g., \citealt{2006A&A...446..395S,2007A&A...475.1159D,2011rich,2007arXiv0712.2898W,2009A&A...494..739S,2010arXiv1008.3143B}). 

We are interested in a related, but more urgent challenge: classification on streaming data, where analysis is performed while the data are still being accumulated. At a logistical level, keeping up with classification (and discovery) assures that the survey producing the data can be continually informed of the progress, allowing the survey to change course midstream if scientifically warranted.  But at a more fundamental level, the reason for real-time classification is that the vast majority of science conducted with time-variable objects, especially one-off transients,  comes when more data are accumulated about the objects of interest. Enabling intelligent follow-up, then, becomes a main driver for rapid classification. Ultimately, one can view {\it classification as a means to maximizing scientific return in a resource-limited environment.}

Given this view of real-time classification, the advantages of a computational (rather than human-centric) approach become clear:
\begin{itemize}
	\item machines, properly trained, are  faster than humans at discovery and classification of individual candidates/events, allowing for operations at arbitrarily high data rates (limited only by computational resources);
	\item the turn-around for well-informed follow-up can be almost instantaneous for computationally based discovery and classification. This allows for more efficient use of the suite of follow-up facilities. For example, observations on a small-aperture telescope can obtain the same signal-to-noise of a fading transient as obtainable on a large-aperture telescope observed after a longer delay;
	\item Experimentation with new discovery and classification schema requires little more than rerunning new codes on existing data, whereas a change to human-based approaches requires additional labor-intensive work with people on a massive scale;
	\item machine-learned classification is reproducible and deterministic, whereas human-based classification is not;
	\item the reproducibility allows for {\it calibration} of the uncertainties of classification probability statements, based on ``ground-truth'' results from the survey itself, with assurances that those classifications are robust as the survey proceeds.
\end{itemize} 
\noindent Robust statements about the demographics of variability of different types requires well-calibrated discovery and classification. And this, in turn, suggests that a machine-based approach is also preferred. Ultimately, there may still be a vital role for humans in the real-time loop, such as serving as ``tie-breakers'' on ambiguous classifications or uncertain follow-up paths for a particular source \citep{ptf10vdl}, but our long-term view is that if a body of human-produced classification statements can be reproduced by machine-learned frameworks, those sorts of statements (during the full-scale production mode of a real synoptic survey) should not come from humans.

In this paper, we describe a methodology and formalism for producing discoveries of astrophysical transients and variable stars using a machine-learned framework based on human expert-trained input (\S \ref{sec:discovery_on_images}). We show how false-negative and false-positive rates can be calibrated with data from the survey itself. In \S \ref{sec:class}, we discuss a machine-based approach to autonomous classification based on {\it feature} sets derived from context and time-series data on individually discovered sources. In \S \ref{sec:ml} we show how a machine-learned model on Palomar Transient Factory (PTF; \citealt{2009PASP..121.1334R,2010SPIE.7735E.122L}) data  produces highly reliable initial classifications\footnote{To be sure, an active group of citizen scientists enabled by the ``Supernova zoo'' also offer an important discovery channel of supernovae \citep{2011MNRAS.412.1309S} within the PTF collaboration that is largely separate from the autonomous discovery and classification framework described herein.}. We end with a discussion about the outstanding challenges and look to future incarnations that may be used on upcoming synoptic surveys.

\section{Discovery on Images}
\label{sec:discovery_on_images}

To identify new sources or brightness changes of known sources in synoptic imaging there are primarily two computational paths: catalog-based searches and imaging-differencing analysis. With the former, sources in each image are found and extracted into a database consisting of flux and position (and associated uncertainty) as well as ancillary metrics on individual detections (such as shape parameters and photometric quality flags). Time-variable sources are then found by cross matching detections on the sky and computing changes in brightness with time. With the latter, a deep {\it reference} image is constructed from several images of a portion of the sky, it is astrometrically aligned with and flux-scaled to an individual image, and it is subtracted from each individual image. The result is a {\it difference image} (e.g., \citealt{2001MNRAS.327..868B}), in which objects are then found and extracted into a database. Since image differencing usually involves the expensive cross-convolution of two images, catalog-based searches are considered computationally faster than image-differencing. Catalog-based searches do well in the regime of large brightness changes and do not suffer from color-correlated misalignment effects due to differential chromatic refraction \citep{2009ApJ...696..870D}. However, in crowded fields (where the typical separation between objects is of order a few PSF distances) or in the presence of high-frequency spatial variations in the background (i.e., near galaxy positions), image-difference searches for variable sources excels\footnote{If all detected objects are to be saved in each epoch, databases derived from image-differencing can be made vastly smaller, since only those sources which change are saved.}. For well-constructed reference images, photometric uncertainties of sources found in image differences can approach the statistical photon limit of an individual image \citep{2000AcA....50..421W}. Given the particular interest in finding variable stars in crowded fields and events  in and around galaxies (supernovae, novae, and circumnuclear sources), especially while the sources are still faint and on the rise, the PTF collaboration chose to perform discovery on image differences. This is also the intended discovery path for most of the new upcoming synoptic surveys: Skymapper \citep{2007PASA...24....1K}, Dark Energy Survey \citep{2005IJMPA..20.3121F}, and the Large Synoptic Survey Telescope (LSST; \citealt{2005AAS...207.2629B,2008arXiv0805.2366I}). The Catalina Real-Time Sky Survey \citep{2009ApJ...696..870D} and the 3$\pi$ survey of Pan-STARRS \citep{2002SPIE.4836..154K} conduct catalog-based searches for transients (cf.~\citealt{2011arXiv1103.5165G}).

 \subsection{Identification}

Frameworks for identifying and characterizing significantly detected objects (e.g., SExtractor; \citealt{1996A&AS..117..393B}) in images can be be applied to image differences. One of the major drawbacks of discovery on image differences, however, is the number of spurious ``candidate'' objects that can arise from improperly reduced new images, edge effects on the reference or new image, misalignment of the images, improper flux scalings, incorrect PSF convolution, CCD array defects, and cosmic rays\footnote{To be sure, some of these effects are also present in catalog-based searches.}. Even with signal-to-noise thresholds and some requirements on metrics related to the candidate shape (e.g., candidate FWHM compared with the image seeing), we have found that the vast majority of SExtracted objects on a given difference image are spurious: in PTF, only about 1 in 1000 \citep{sahand} extracted candidate objects (considered to be at least as significant as a 5-$\sigma$ detection) in a typical field are what we would deem to be astrophysically ``real'' (i.e., an origin owning to a change beyond the Earth's atmosphere). Nugent et al.\ (2011) provides details on the SExtractor extraction requirements and which candidate/subtraction parameters are saved into a database.

 \subsubsection{Real or Bogus?}
\label{sec:rb}

Beyond the subtraction and source extraction steps, our first significant challenge is in determining which of the candidates are worth pursuing as real astrophysical events and which are ``bogus.'' With training, many astronomers can identify when
subtractions are poor or if a candidate is dubious to reasonable
accuracy. But given the rate of candidate extractions, about 1--1.5 million per night for PTF, it is clearly not feasible to present candidates to human scanners to determine the reality of every candidate. To keep data volumes small enough to be human-scanned, several options are available. First, restrict the candidates to a certain domain-specific set. For example,  scanning only those candidates that are near but offset from extended galaxies will generally succeed at finding some supernovae (and ignore most variable stars), but it will fail to find supernovae far from their host galaxy, supernova associated with low-luminosity hosts, and supernovae near the centers of galaxies (cf., \citealt{2011ApJ...732..118S}). There are active areas of research in all three of these cases (e.g., \citealt{2010AJ....139.2218M}). Second, require several candidates to appear at or near the same location on several epochs. This is indeed good at mitigating against cosmic rays and other transient artifacts, but mis-subtractions tend to correlate at the same locations even at different epochs (that is, when a subtraction is bad at some position on the sky at some epoch, there is an increased tendency for it to be bad at other epochs). This approach also runs the risk of waiting until too late to identify a (short-lived) astrophysical transient. Third, impose restrictive threshold cuts on the derived parameters of the candidate and the subtraction, such as requiring a 30-$\sigma$ detection with a shape that is well-fit by the inferred PSF of the image. But, since most (real) candidates occur near the detection threshold and there is no guarantee that highly significant flux differences are all due to real astrophysical events, this approach will systematically exclude the lion's share of real events.

Our approach---to remove the human element in any real-time decision processes---is to use machine learning to provide a statistical  statement about whether a given candidate should be considered astrophysically real or spuriously bogus. Such statements can then be combined over several epochs, if required, to determine if that identified candidate should be considered a {\it discovery} of an astrophysical source. To arrive at deterministic statements about each candidate, there are three broad classes of inputs that can be used to create a  ``labelled'' set of candidates for use in the machine training: use trained/expert human scanners to opine on the real/bogus nature of a subset of the candidates, add a set of artificial sources to the raw data, or construct a ground-truth labelled set by using knowledge of which candidates turned out to real based on follow-up observations (e.g., using spectroscopically identified supernovae) of earlier incarnations of the survey. 

\begin{figure*}[tbp]
	\centerline{\includegraphics[width=5in]{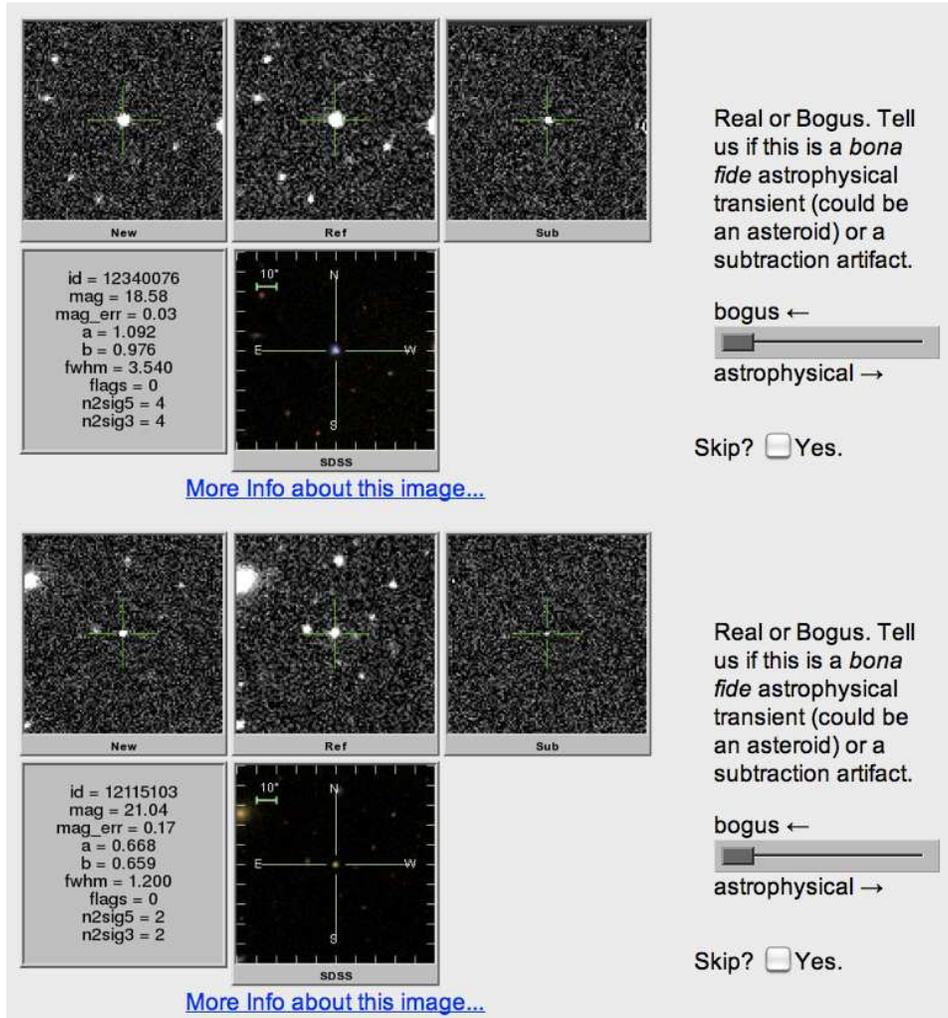}}
\caption[rb]{Example webpage showing two subtractions (1 $\times$ 1 arcminute thumbnails of the deep reference image, the new image, and the subtraction image, left to right; the bottom panel shows the SDSS image) presented to human scanners; some metrics of the subtraction (such as the FWHM of the candidate source) are shown to the user to help them make a decision with more than just visual information. The responses were generated by a slidebar indicating the scanner's thoughts on whether the subtraction was bogus or astrophysical.}
\label{fig:rb1}
\end{figure*}

Each labeling approach inheres advantages and drawbacks:
\begin{itemize} 
	\item {\bf Human-scanned}: Having humans provide the labels can ensure, by construction, that the machine-trained statements closely mimic what someone looking at a certain candidate might say about it. To fully capture the broad range of astrophysically real or spuriously bogus candidates, however, many (perhaps thousands) of candidates must be tediously labelled by hand. Moreover, there is no guarantee that a real source (especially near the detection threshold) will be labelled as such; and the converse is also true: bogus candidates might be spuriously labeled as real even by experts. 
	\item {\bf Artificial-source constructed}: Though computationally intensive, ``fake'' events can be placed at a variety of locations on the sky: in regions of high stellar density, near CCD chip edges, and at a variety of locations around a large diversity of galaxies. The main difficultly is in ensuring that the artificial candidates inserted into raw data are a close-enough representation of what a real source would look like in each image. That is, if all relevant  effects (of the atmosphere, camera optics, telescope shake, etc.) are not properly modeled then there is a risk of a mismatch between what the derived parameters of the fake sources are and how real events are manifest in that parameter space. 
	
	\item {\bf Ground-truth derived}: A ground-truth construction benefits from explicitly removing the vagueness and non-repeatability of human scanning but, in some cases, there remains an implicit reliance on human labels. For example, if spectroscopically identified supernovae are used to construct the ``real'' label set then there is a built-in bias towards spatial configurations that led previous observers to decide to follow-up such events. Further, if a catalog of known variable stars is used then there is bound to be a mismatch in survey characteristics; only bright variable stars, for instance, might be labelled as real. Determining bogus labels directly is difficult.
\end{itemize}
\indent As there is no pure labeling process, we initially chose to use the human-scanned approach for the PTF data. (\citealt{sahand} describes new efforts centered around the ground-truth approach). To facilitate the human labeling, we built a web-based system, called ``Group/think'' based on the Python computing language\footnote{\url{http://python.org}} and the Google App Engine framework (\citealt{1507544}; Figure \ref{fig:rb1}). During the commissioning phase of the project, several of the PTF collaboration members who had been hand-scanning each candidate every night were presented a series of images (each showing the reference image, the new image, and the subtraction) and asked to determine if the subtraction was ``bogus'' or ``real,'' allowing them to assign a confidence level to their choice ranging from 0 (definitely bogus) to 1 (definitely real). The initial catalog, all based on R-band filter data, consisted of 370 candidates chosen from the first few spectroscopically confirmed supernovae discovered in the PTF commissioning ($\sim$15\% presumably real) and a set of the nearest ($\sim$85\%  presumably spurious) candidates to those supernovae. These candidates tended to be either obviously real or obvious bogus. A ``realbogus'' classifier was trained on those labels (see \S \ref{sec:rb}) and applied to the first month of commissioning data. From that data, we created a new set of 574 candidates which spanned the range bogus to real, with a concentration of candidates intermediate to the two extremes. 

\begin{figure*}[tbp]
	\centerline{\includegraphics[width=5.5in]{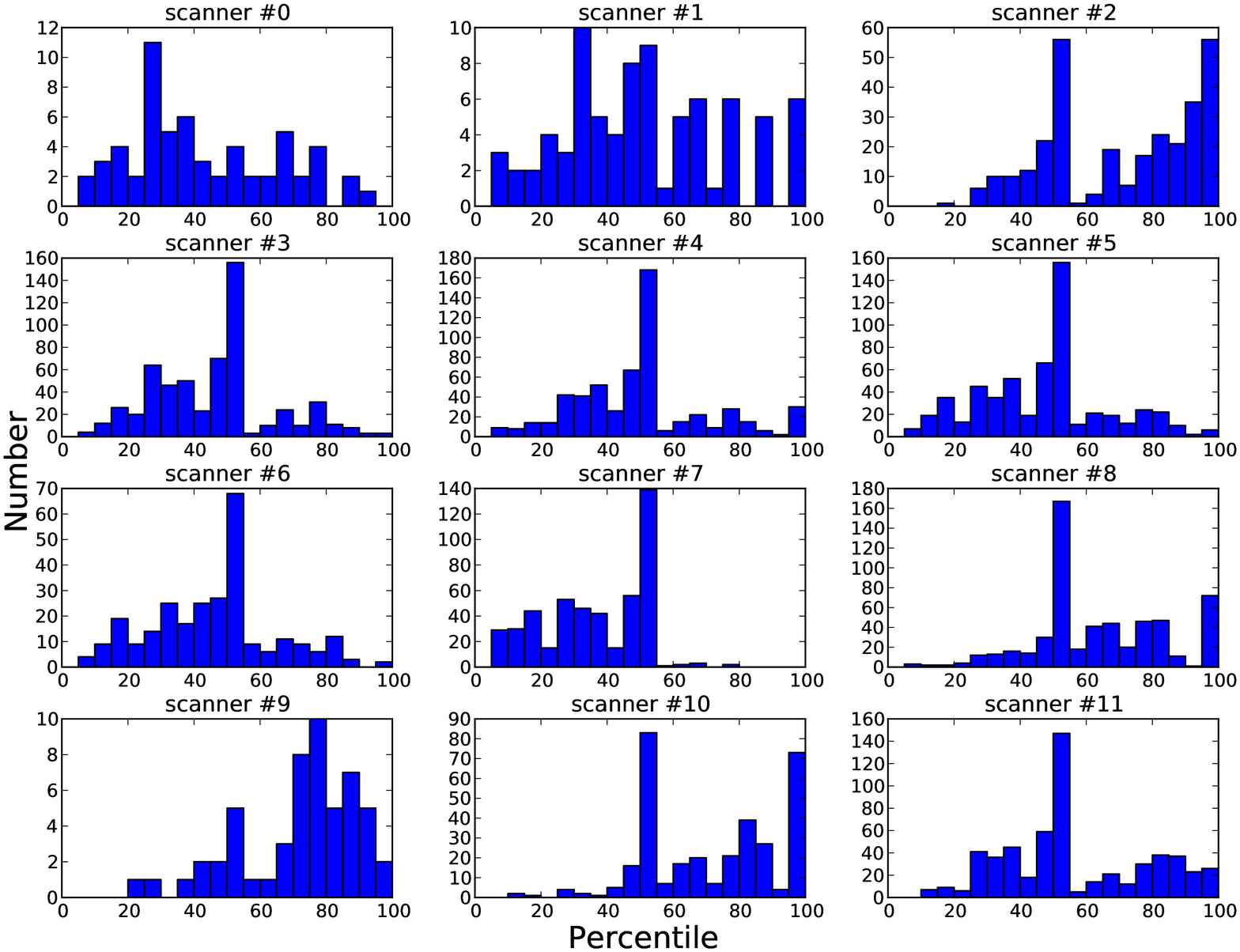}}
	\vspace{1cm}
	\centerline{\includegraphics[width=3in,angle=90]{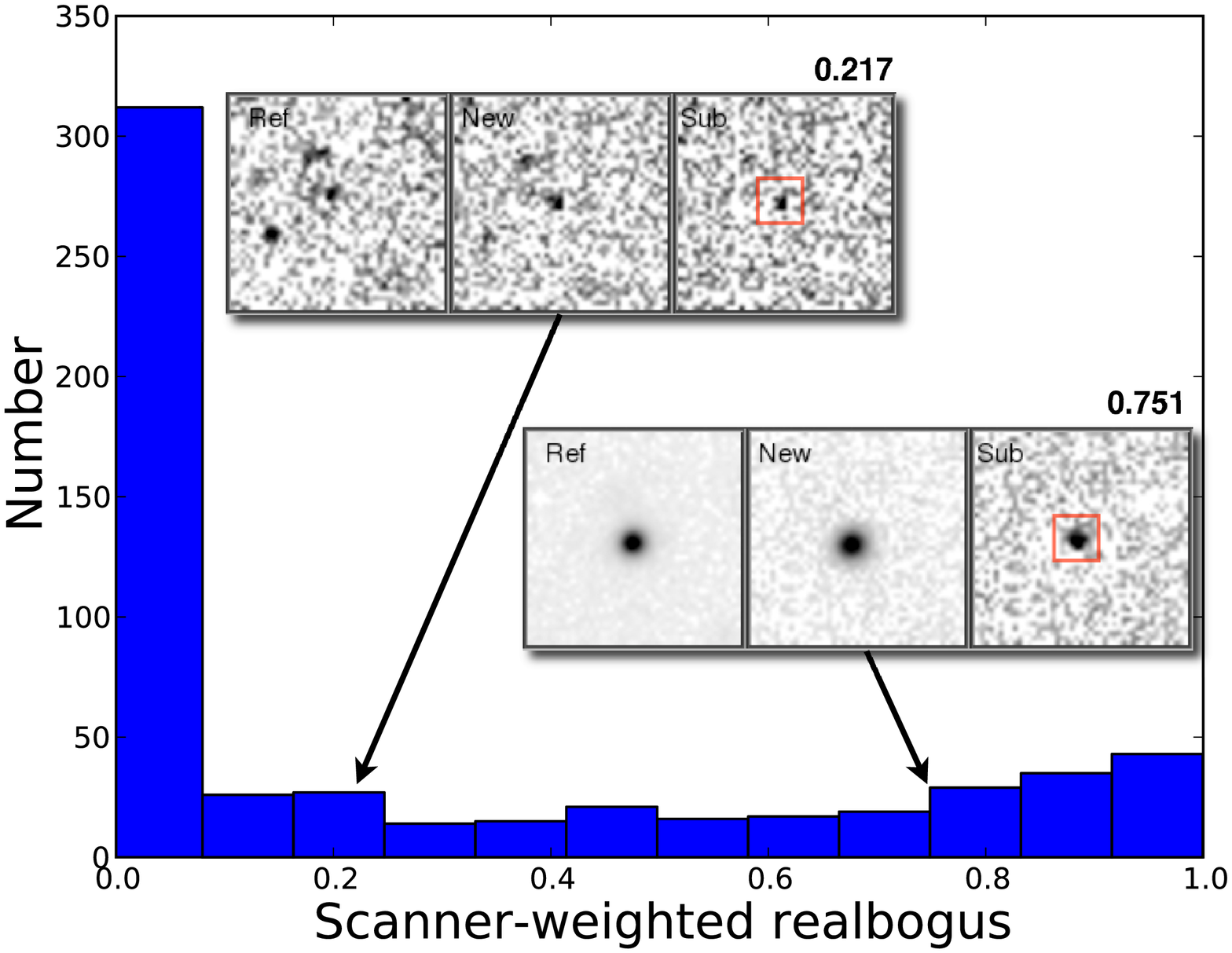}}
\caption[rb]{(top) Distribution of training set scoring for 12 human scanners over 574 subtraction candidates. These distributions were used to compute the weights and biases for each scanner. (bottom) Final scanner-weighted distribution of the training realbogus set. Examples of probable (0.22) and likely (0.75) subtraction candidates are shown.}
\label{fig:scan}
\end{figure*}

So as not to bias the labeling to any one scanner, we determined the bias of each scanner relative to the group of scanners. Figure \ref{fig:scan} shows the percentile distribution for each scanner relative to the other scanners for each candidate that that scanner marked up. If all scanners for a given candidate gave the same realbogus value, then we assigned 50 percentile to every scanner. While most candidates show broad agreement, it is clear that some scanners were more or less optimistic in the aggregate than the group. Scanners \#5--7 appear to believe fewer candidates are real and scanner \#2 was more optimistic.
For a given scanner, their bias is determined from a mean of the percentile ranks of all their scanned candidates and an estimate of their 68\% confidence scatter is determined using a Bayesian estimate, assuming a Jefferys prior \citep{Jeffreys24091946} for the standard deviation. Larger scatter indicates that the scanner agrees less often with the group. For every candidate, we create a realization of the debiased score for each scanner adding it to a temporary list if that scanner's standard deviation (std) of percentile is less than a number chosen randomly from 0 to 100. Since the typical values of std range from 15 to 25, approximately 80\% of a scanners biased score is used in a given realization. We take the median of 50 realizations of such lists. In this way, we create a scanner-weighted realbogus score for our labelled training sets.  Figure \ref{fig:scan} shows the distribution of the scanner-weighted realbogus score for the second labeling run of 574 candidates.

We wish to construct a parameter---generated rapidly at the time that the image differencing is completed---which reasonably mimics the human scanning decision of real or bogus. This necessitates the use of readily available metrics from our subtraction database on the candidates themselves used as input to train a machine-learned (ML) classifier (as opposed to some metrics which might be gleaned from external databases). For each candidate in the training sample, we derived 28 metrics (called {\it features} in ML parlance) from the SExtractor output (Table \ref{tab:features}).  Ill-derived (e.g., division by zero) or absent features were considered ``missing'' data for the purposes of the learning process. The scanner-weighted  realbogus score for the training set was used as the ground-truth label for each candidate.

We explored ML-regression to predict the realbogus numerical value but found available techniques  ill-suited to handle missing data and data with a mixture of numeric and nominal features. Instead, we created 5 nominal classes based upon the numeric scanner-weighted  realbogus label: \verb bogus ~($< 0.10$), \verb suspect ~([0.10,0.40)), \verb unclear ~([0.40,0.70)), \verb maybe ~([0.70,0.95)), \verb realish ~($\ge$0.95). Using the WEKA framework \citep{weka}, we trained a random forest classifier \citep{2001brei}, using 10-fold cross validation, on the labelled data and developed a ``cost'' matrix to penalize gross misclassifications and to mitigate the effect of having many more bogus candidates than reals in the training sample. The classifier produces a probability $P_i(C_j)$ of the $i$-th candidate belonging to each of the $j=5$ classes. The ML-trained realbogus value for the $i$-th candidate is constructed using:
\begin{equation}
RB_i = \sum_j P_i(C_j) \times w_j,
\label{eq:rbi}
\end{equation}
where the class weights for $C_j$ = [\verb bogus , \verb suspect , \verb unclear , \verb maybe , \verb realish ] were set, ad hoc, to be $w_j = [0.0, 0.15, 0.25, 0.50, 1.0]$.

\begin{figure*}[tbp]
	\centerline{\includegraphics[width=4.5in]{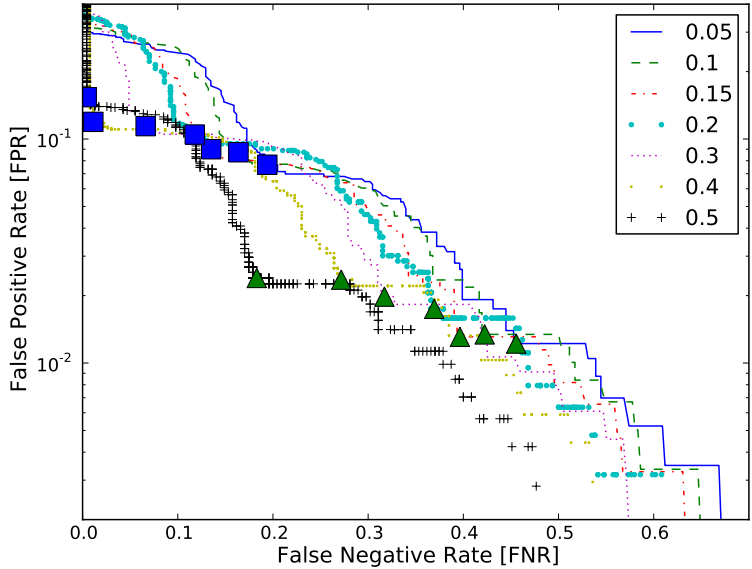}}
\caption[rb]{ROC Curve for the trained realbogus sample as implemented for the Palomar Transient Factory. The seven curves were generated using a cut on the scanner-weighted RB scores (value shown in the legend) where all candidates with that cut value or larger were assumed to be definitely real and those below definitely bogus. The higher the value, the more conservative the human discovery threshold would be. Those candidates that were ``real'' but below the ML-determined realbogus cut value (for several cuts) were considered false negative (Type II error). Those candidates that were ``bogus'' but above the ML-determined realbogus cut were considered false positive (Type I error). The blue squares (green triangles) show the results for each curve assuming an 0.2 (0.4) ML-determined realbogus cut. }
\label{fig:rbroc}
\end{figure*}

To evaluate the effectiveness of the classifier we constructed a ``receiver operating characteristic'' (ROC) curve (Figure \ref{fig:rbroc}) showing the false-negative rate (FNR; real candidates set as bogus) versus the false-positive rate (FPR; bogus candidates selected as real) for a variety of different real/bogus cuts on the training data and the learned results. At an ML-determined realbogus cut of 0.2 we expect an FPR between $0.08$--0.12 and a false negative rate between $\sim$0.02--0.2 (the range of uncertainty comes from an uncertainty in where the true  cut should be for the scanner-weighted realbogus). At an ML-determined realbogus cut of 0.4 we expect an FPR between $0.01$--0.02 and a false negative rate between 0.18--0.45. In \S \ref{sec:dis}, we discuss how these ROC curves are used in the discovery process.

\begin{figure*}[tbp]
	\centerline{\includegraphics[width=4.5in]{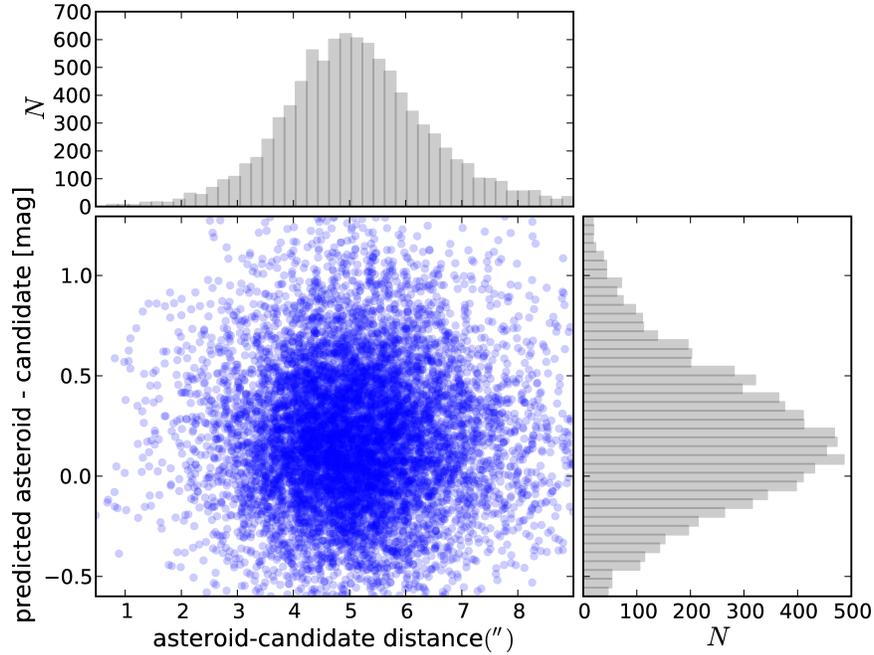}}
\caption[rb]{Distribution of the offset and magnitude differences of asteroids in the validation sample from the nearest-detected candidate in the PTF subtraction database. There is a clear locus of candidates from 2 to 8 arcsecond of the predicted position and within $\sim$1.5\,mag of the predicted brightness at the time of observation. The overall magnitude difference and scatter is expected given the PTF filter plus zeropointing uncertainties coupled with the approximate nature of Minor planet magnitude predictions. The positional offset is likely due to a combination of imprecise absolute astrometry on PTF images (improved since 2009) as well as the approximate nature of the orbit calculations in PyMPC: the code  makes use of the orbital parameters downloaded from the Minor Planet Center that are updated only monthly, and do not include the most precise small-body gravitational perturbations. For the purposes of the creation of this validation set, the positional offsets are not important.}
\label{fig:rocks}
\end{figure*}

To validate the ML-classification we created a list of known asteroids passing through 4150 subtractions (=2615 deg$^2$) over three nights of data in Fall 2009 (starting at UJD = 2455045.6648). These data should be fairly typical, representing the diversity of fields and image quality in the survey: observations during these nights were not biased towards or away from the stagnant asteroid zone nor were they especially focused on imaging in the Galactic plane. The catalog positions and calculated magnitude of each asteroid were found for each subtraction, using a custom parallelized Python code that made use of the Minor Planet Center asteroid data tables\footnote{\url{http://minorplanetcenter.net/iau/mpc.html}}. This code, which typically runs 10$\times$ to 100$\times$ faster than queries to the minor planet center site, is made available by us for the community as an open webservice\footnote{\url{http://dotastro.org/PyMPC/webservice\_readme.html}.}.  We identified 19954 asteroids within the subtraction footprint. We created a subsample of those with good ($<10-15$ \arcsec) a priori location accuracy from the catalog, bright enough to have been detected (i.e., the catalog magnitude at least as bright as the limiting magnitude of the image), and which were not close to the edges of the arrays (position $>30 \arcsec$ from the nearest edge). Further, so as not to identify candidates associated with elongated asteroid observations, we restricted the sample to asteroids calculated to have a proper motion at the time of observation of less than 50 arcsecond per hour, resulting in less than 0.83 arcsecond of total motion during the one minute exposure. There were 9034 asteroid-associated candidates in this subsample. Figure \ref{fig:rocks} shows the distribution of asteroids relative to the nearest candidates on the sky.

\begin{figure*}[tbp]
	\centerline{\includegraphics[width=5.5in]{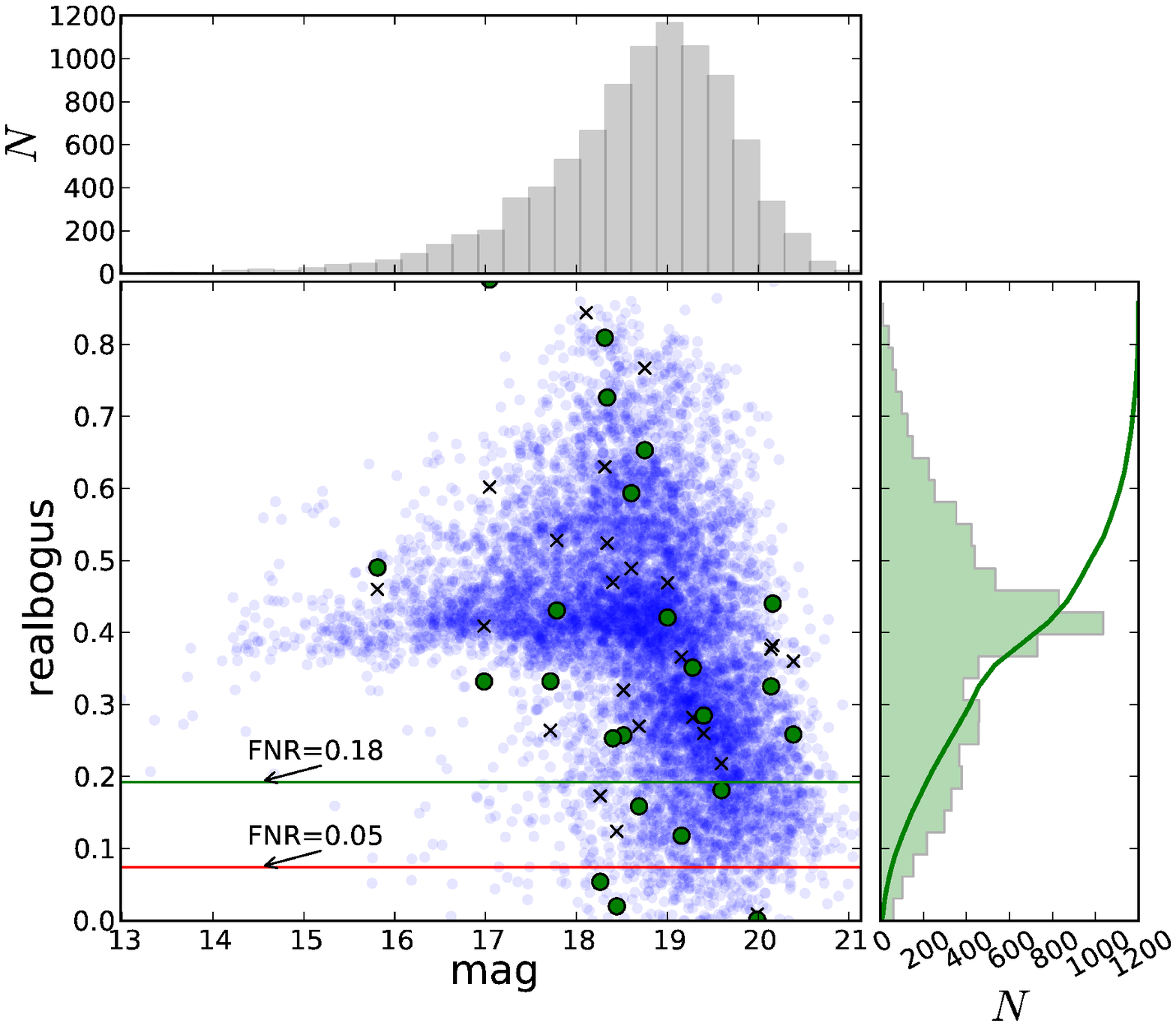}}
\caption[rb]{Distribution of asteroid-associated candidate ML-classified realbogus values. The cumulative distribution of realbogus values is shown as a green curve in the outset histogram at right. The horizontal lines show two effective false-negative rates (misidentified real candidates) for two different realbogus cuts. Crosses note the 24 asteroids in the subsample that are within one arcsecond of a source detected in the reference image. Green circles show the contextualized realbogus ``score'' for those candidates (\S \ref{sec:score}).}
\label{fig:rocksrb}
\end{figure*}

Figure \ref{fig:rocksrb} shows a validation of the ML-classified realbogus on these candidates. Nominally all these candidates are taken to be bona fide ``sources,'' providing a ground-truth set for us to test the ML-classifier. [In practice, however, near the faint end of the distribution there will be some pollution of this set with bad-subtraction candidates: if a known (faint) asteroid happens to be near a poorly-subtracted region, that candidate will be incorrectly included in the sample.] There is a clear trend for brighter candidates to receive a higher realbogus value. There are many  sources with realbogus around 0.35--0.50 that show no trend with magnitude; this locus reflects the distribution of the classifier output convolved with the weighting scheme (eq.~\ref{eq:rbi}). The line near realbogus=0.2 (FNR=0.18) is in rough agreement with, but higher than, the FNR predicted ($\sim 0.11$) from the training set (Fig.\ \ref{fig:rbroc}; blue squares).  This difference might be in part explained by the inclusion of bad-subtraction candidates in the asteroid set.  Since most asteroids are found far from stars and galaxies on the sky, there is a legitimate concern that this introspection is only validating the ability of the ML-classifier to identify spatially-isolated transients. However, by selecting the two dozen candidate asteroids that happen to be near ($<1 \arcsec$) detected objects in the reference images ($\times$ symbols in Fig.\ \ref{fig:rocksrb}) we find no clear trend of those sources to be preferentially different in their realbogus values.

 \subsubsection{Contextualized Statements}
\label{sec:score}

The metrics used in automatically classifying individual subtractions (Table \ref{tab:features}) relate entirely to the candidate itself and not its surroundings (save the {\tt good\_cand\_density} parameter). Candidates generated from poor subtractions---often owing to mis-registration and/or to a poorly characterized convolution kernel---tend to cluster spatially. A high realbogus value on one candidate might be considered suspect if neighboring realbogus values are also high (under the reasonable assumption that significant variability is not common and should not be spatially correlated); the most egregious example would be when the misalignment of the new and reference images are more than a few times the scale of the seeing, leading most candidates 
on that subtraction to have high realbogus values.

This consideration calls for a contextualized statement of realbogus that takes into account what has happened both locally and globally on the subtraction. A simple scaled realbogus value is determined in the PTF pipeline by taking the ratio of the candidate realbogus to the mean of the nearest two candidates' realbogus values on that subtraction. A more complex scaled realbogus value takes into account all candidates in the subtraction frame, weighting more heavily those other candidates nearby to (and with similar magnitudes of) the candidate and the reference source themselves. We create a contextualized score with an ad hoc formula that takes into account the realbogus value itself and the two scaled versions. The score serves to downweight the candidates whose realbogus is not much higher than neighboring realbogus values. The score is also downweighted for sources very near diffraction spikes or bleeding trails near very bright stars (mag $< $ 13). In PTF, scores are used to rank-order discoveries from most promising to least likely.

\subsection{Discovery}
\label{sec:dis}

If the unit of discovery---the moment of identifying an event as a true astrophysical source---was only a realbogus statement (and associated score) about a single candidate, there would be enormous inefficiencies and impurities in PTF. Roughly 10\% of all bogus candidates would be ``discovered'' and $\sim$20\% of all real sources would be missed (Figures \ref{fig:rbroc} and \ref{fig:rocksrb}). Moreover, at the single-epoch sensitivity limit of PTF, there are at least as many as 10 times the number of slow-moving asteroids as stationary transients and variable stars, meaning most discoveries would be of asteroids and not the events and variables of interest. To mitigate against asteroid detection, PTF is generally scheduled to observe away from the stagnant asteroid zone and, more importantly, places a high priority in getting more than one image of the same field in a given night separated by at least 45 min--1 hour \citep{2009PASP..121.1395L,2009PASP..121.1334R}. By requiring two reasonably good candidates to be coincident in space but separated by at least 45 minutes in time, we largely avoid asteroid ``discovery'' and can also build a higher degree of confidence in the astrophysical nature of the variability.

Since multiple candidates are required for discovery, the ROC curves for a single candidate are not the appropriate measure of efficiency and purity (${\cal P}$) of discovery. We define purity as
\begin{equation}
{\cal P} = \frac{\cal R_{\rm dis}({\rm real})}{\cal R_{\rm dis}({\rm real}) + \cal R_{\rm dis}({\rm bogus})}
\label{eq:s}
\end{equation}
where the rate of discovery of real sources is: 
$${\cal R_{\rm dis}({\rm real})} = {\cal R({\rm real})}\, P({\rm discovery}|{\rm real})$$
and the rate of discovery of bogus sources is:
$${\cal R_{\rm dis}({\rm bogus})} = {\cal R({\rm bogus})}\, P({\rm discovery}|{\rm bogus})$$
Note that $P({\rm discovery}|{\rm real})$ is just the {\it efficiency} of discovery. We expect in PTF (and other imaging surveys where detections are made on subtractions) that in any single subtraction $R({\rm bogus}) \gg R({\rm real})$. Roughly, in PTF, $R({\rm bogus}) \approx 1000 \times R({\rm real})$. If discovery were done on just a single epoch then, 
$P({\rm discovery}|{\rm real}) = (1 - FNR)$ and $P({\rm discovery}|{\rm bogus}) = FPR$.  Following \S \ref{sec:rb}, with $FNR \approx$ 0.2 and $FPR \approx 0.1$ this implies ${\cal P} = 0.008$; this is unacceptably low. If we adopt a more conservative cut (Fig \ref{fig:rbroc}), with $FNR \approx$ 0.4 and $FPR \approx 0.01$, then ${\cal P} = 0.06$.

To keep ${\cal P}$ near unity (a high purity of discoveries to maximize followup resources), equation \ref{eq:s} requires that we create a detection classification scheme that satisfies $P({\rm discovery}|{\rm real}) \gg P({\rm discovery}|{\rm bogus})$.   When multiple detections are required to cross a threshold for a discovery, then $P({\rm discovery})$ changes, and importantly, this probability changes differently for bogus events than real events. In the simple case where two observations are made and two good detections (i.e., high realbogus) required, then
$$P({\rm discovery}|{\rm real}) = P({\rm good~detection}|{\rm real}) \land P({\rm good~detection}|{\rm real}) = (1 - FNR)^2.$$
This assumes that the probability of getting the same classification value is the same  for both epochs, which might be nominally expected in the case of a source with approximately constant flux and similar observing conditions. For a bogus source to be called a discovery, however, two bogus subtraction candidates must be both incorrectly identified as real and occur close on the sky. In PTF, we have found that the existence of a bogus candidate is (unfortunately) highly correlated with the existence of another bogus candidate near the same place on the sky at different times: that is, certain places on the sky will preferentially yield bad subtractions (due to a combination of poor astrometry, imperfections in the reference image, and proximity to bright stars or chip edges). Ignoring correlations of realbogus values\footnote{The positive correlation between bogus detections means that $P({\rm 2nd~detection~|~bogus,1~detection}) > P({\rm detection|bogus})$,  implying that $P({\rm discovery|bogus}) = P({\rm 1st~detection|bogus}) \times P({\rm 2nd~detection~|~bogus,1~detection})  > FPR^2$.}, we expect with the two-candidate requirement ${\cal P}=0.06$ and ${\cal P}=0.78$ for $FNR = 0.2$ ($FPR =0.1$) and $FNR = 0.4$ ($FPR =0.01$), respectively. This means  for the 2-candidate discovery process that at 78\% purity, we are 36\% efficient in finding real sources.

In practice, the source-discovery process in PTF is complicated by the fact that real source brightnesses are changing in time (and so too the respective realbogus values). We were also wary of missing faint (and low realbogus-valued) events occurring in nearby galaxies and so decided to err on the side of lower purity and higher efficiency. Since much of the science of PTF is focused on fast variables and short-lived transients \citep{2009PASP..121.1334R}, we also search for sources that are changing on relatively short timescales. Indeed, in the current incarnation of the framework, our initial query of the candidate database returns all candidates in a certain date range with realbogus greater than 0.17 (and with contextual realbogus greater than 3.3 times nearby sources;  \S \ref{sec:score}). The positions of these candidates are then cross-matched with other candidates with realbogus $\ge 0.07$ within 2.0 arcseconds on the sky that were imaged at least 45 minutes (and no more than 6 days) before or after the candidate. PTF tries to obtain at least two epochs on the same part of the sky per night and repeat visits that part of the sky every 3--5 days. Given this we are reasonably assured that, if the source is real and still detected, at least one other candidate will be matched given the temporal criteria\footnote{Clearly, when weather adversely affects observing over several nights real sources may go undiscovered because of this temporal windowing.}. As a fail-safe against missing bright nearby supernovae, human-scanners are presented candidates near large resolved galaxies with a much lower realbogus threshold \citep{2011MNRAS.412.1309S}.

Once a set of subtractions have finished loading (typically every 45 minutes for the  10$^5$ candidates in 100--200 square degrees of imaging) into the real-time subtractions database housed at Lawrence Berkeley National Laboratory (LBNL), an email with the date range of the subtractions is sent to an account which is then parsed automatically by a script running on the University of California, Berkeley campus. Depending on the density of stars in the field and the prior cadences in that part of the sky, typically 30--150 sources are identified. These sources, and associated candidates, are then saved as preliminary discoveries in an internal database of the automated system. The $\sim$10$^5$ candidates generated per reduction run are typically vetted in 5 minutes via remote database queries.

\section{Classification}
\label{sec:class}

Discovery inheres no more insight than the identification of a set of candidate events as belonging to a changing astrophysical source. The physical origin of the emission---the classification of the source into an established hierarchy of known variable and transient types---requires a different set of questions and another round of inspection now abstracted from the 2-D images. Indeed, once a source is preliminarily discovered, classification is done using only the data derivable from the LNBL databases and other (remote) webservice queries.

The PTF collaboration maintains a database of source discoveries, each assigned a unique name (such as PTF 09dov). During commissioning and during the start of the science operations of PTF, sources were discovered by human scanners who looked at individual candidates and associated candidates at other epochs. ``Discovering,'' in that context, required that a button  be clicked on a candidate scanning webpage. At the time of discovery, the scanner is also asked to suggest a crude classification choice, between variable star ({\tt VarStar}), transient ({\tt Transient}), and asteroid ({\tt rock}).  To mimic this interaction, removing the need for human scanning, one of the main roles of the automation is to provide the same set of initial classifications based on available data. As we now describe, the classification routines also try to provide more refined statements about the nature of the variability.

\subsection{Features based on available data}

At a given place in the sky, there are broadly two categories of information available (in principle): the changes of brightness in time as a function of wavelength and the context of where a source is located in relation to known objects (e.g., stars and galaxies) and coordinates (super-galactic plane, ecliptic, etc.). Context information also includes the metrics on those nearby objects, such as color, apparent size, redshift, and spectroscopic type. To condense and homogenize all of the available information on a given transient or variable, like with image classification, we compute both context and time-domain features which may be used in decision rules or in a machine-learned classifier.

Since one the primary goals of the PTF collaboration is to rapidly identify new transient sources or extreme variable stars (e.g., \citealt{ptf10vdl}), we wanted to build a classification engine that was capable of making decisions with only a few epochs of imaging. To this end, we generated time-domain features that could have meaning in the limit of even a small number of epochs\footnote{There is a rich and growing literature that makes of many epochs of high-quality photometry to produce robust classifications on variable stars, quasars, and supernova. See \citet{2011arXiv1104.3142B} for review.}. Those features are described in Table \ref{tab:tdfeatures}. 

\subsubsection{Context}
\label{sec:context}

With limited time-domain data available, it is clear that strong classification statements can be made based on context alone. A variable point source with quiescent colors in the SDSS bands of $0.7 < u - g < 1.35$ mag and $-0.15 < g - r < 0.4$ mag is very likely an RR Lyrae star \citep{2010ApJ...708..717S}. A transient source near the outskirts of an intrinsically red galaxy is very likely a type Ia supernova.  When a new discovery is made, in addition to computing the time-domain features, we  make separate HTTP/GET external database queries to SDSS (DR7), USNO-B1.0, and SIMBAD. We also search a database of galaxies within 200 Mpc and record the projected offset of the source to the nearest galaxy. For all queries, information about nearby sources (and the distances to them) is saved in a database and associated with the newly discovered source. A subset of that information is converted into features for that source and becomes available to the classifier. Table \ref{tab:contextfeatures} describes our context features. Some of the features are determined ad hoc (such as {\tt usno\_host\_type}) based on experience with these catalogs. In a few cases, where the position is nearest (but not consistent with) the position of a star and consistent with a large SDSS galaxy, we will assign that galaxy as the host.   In addition to {\tt usno\_host\_type}, we also make a complex decision about the best ``host'' type using the SDSS and the local galaxy catalog. In particular, if {\tt SpecObjAll.specClass} is ``galaxy'' or {\tt near\_local\_gal} is ``yes'' or {\tt apparently\_circumnuclear} is ``yes'' then we set {\tt best\_host\_galaxy} to ``galaxy.'' If {\tt SpecObjAll.specClass} is ``qso'' and the {\tt sdss\_spec\_warning} does not contain ``NOT\_QSO'' then we set {\tt best\_host\_galaxy} to ``qso.'' We set {\tt best\_host\_galaxy} to ``star'' otherwise.

\subsection{Oarical}

\begin{figure*}[p]
\centerline{\includegraphics[width=5.5in,angle=0]{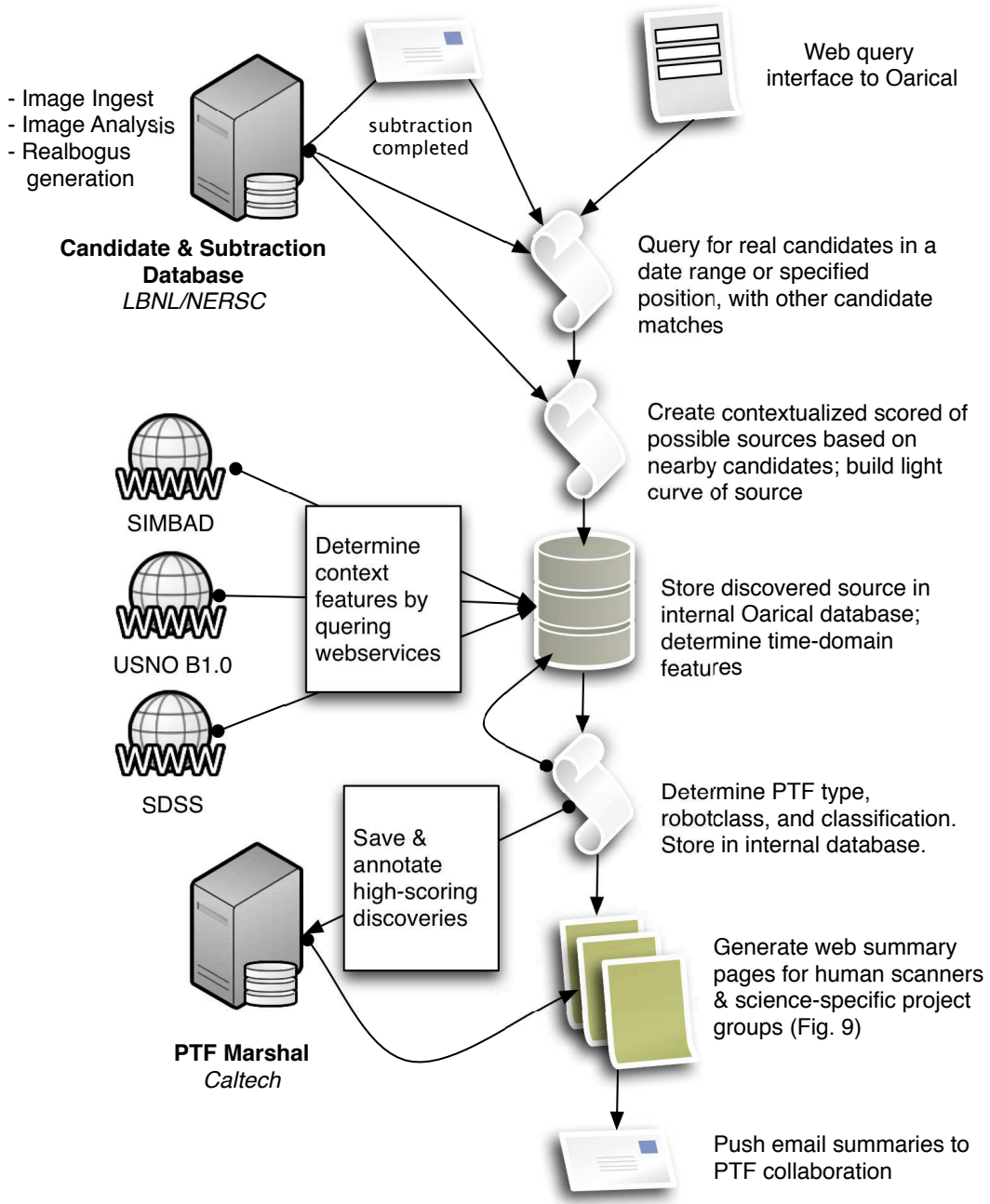}}
\caption[rb]{Flow diagram of Oarical, showing the major input and output components of the classification framework for PTF.}
\label{fig:flow}
\end{figure*}

The main purpose of the classifier, which we call Oarical,
is to quickly label a newly discovered source with as much specificity as possible and with as little time-series data as available. In particular, since the main science of the PTF collaboration focuses on transient/explosive events on short timescales, a particular premium was placed on the ability to recognize such events (i.e., supernovae, extragalactic ``gap transients'', novae, and galactic outbursts). The workflow and major interfaces are diagrammed in Figure \ref{fig:flow}. The heavy reliance on context features (\ref{sec:context}) reflects the immediacy of the transient classification.  The initial classification (Figure \ref{fig:taxonomy}) is separated into four groups: {\tt VarStar} (variable star), {\tt SN/Nova} (supernova or nova), {\tt AGN-cnSN-TDE} (circumnuclear event, such as a tidal disruption flare, AGN/QSO activity or a circumnuclear supernova), and {\tt rock} (asteroid). We produce an ordering of confidence of each classification for all discovered sources (what the most likely class is) and an overall scale of the confidence in the most likely class. If the discovery score of the source itself is low (near the realbogus discovery threshold) that scale will be low as well.

\begin{figure*}[tbp]
	\centerline{\includegraphics[width=5.5in,angle=0]{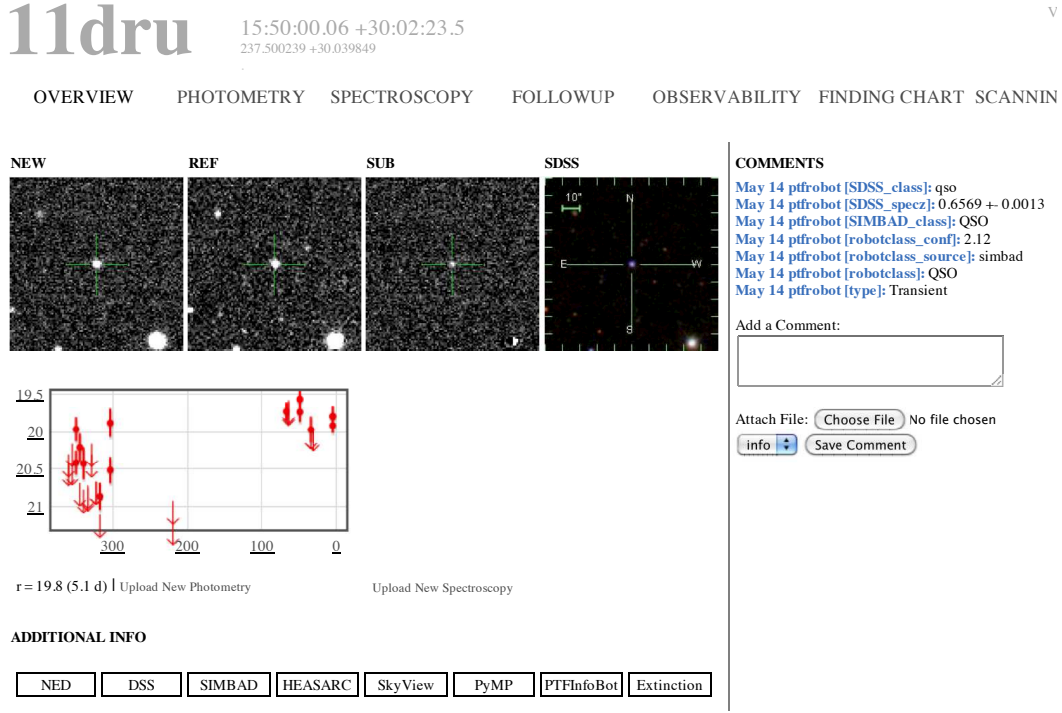}}
\caption[rb]{Screenshot of the PTF Marshal with automatic annotations from Oarical (``{\sc PTFROBOT}'').}
\label{fig:ann}
\end{figure*}

Oarical started routine operations on April 6, 2010 with the first robotic discovery and classification of PTF\,10fhb ($\alpha$(J2000): 10\fh17\fm00\fs.30, $\delta$(J2000): +45\arcdeg30\arcmin48\arcsec.2). Spectroscopic followup of PTF 10fhb with the Double Spectrograph on the Palomar 200 inch Telescope on 12 April 2010 revealed it to be a Type Ia supernova near maximum light at redshift $z=0.1329$. During each night, after each subtraction run has completed (usually every 30--45 minutes), Oarical operates on the candidates from that subtraction run with discoveries noted in an internal database (following \S \ref{sec:dis}); high-scoring sources are saved automatically as PTF-named events in the ``PTF Marshal.'' The PTF Marshal is a database housed at the California Institute of Technology (Caltech), which serves as the official central repository for discoveries, followup, and collaboration interaction over PTF sources (see Fig.~\ref{fig:ann}). Initial classification, following the prescription below, are also saved into the PTF marshal.  Given the complex decision process used by the PTF collaboration in determining which discovered sources are followed-up spectroscopically \citep{ptf10vdl}, we do not have an unbiased view of the success rate and error rates in the Oarical classification (see below). 

During the first year of the PTF survey, one of the main challenges in getting Oarical to produce reliable classifications was the lack of sufficient PTF data and  ground-truth sources to train a machine-based classifier (we discuss this further in \S \ref{sec:disandconc}). As such we built and refined a tree-based classifier to match our own {\it expectations} of classification based on a series of decisions using the context and time-domain features. In this classifier, we rely on a hierarchy of input authorities, from most reliable to less reliable:
\begin{figure*}[tbp]
	\centerline{\includegraphics[width=5.5in,angle=0]{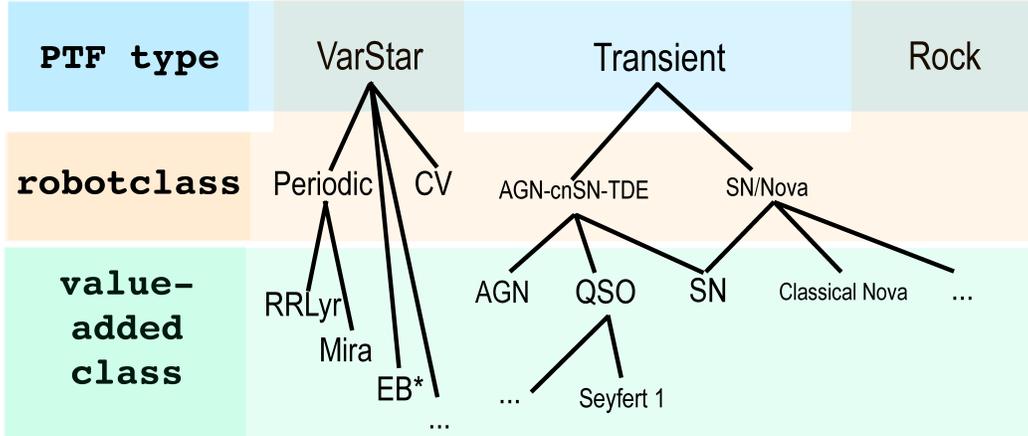}}
\caption[rb]{Taxonomy of classification used by Oarical. The top bar shows the PTF type, the initial classification used when saving candidates as sources. The second tier, ``robotclass,'' shows the four classifications determined by Oarical for a new source. The bottom tier shows example classifications determined from SIMBAD identifications and SDSS spectroscopic analysis.}
\label{fig:taxonomy}
\end{figure*}
\begin{enumerate}
  \item {\it Minor-Planet Center:} After the context and time-domain features are assembled, Oarical queries our parallelized minor-planet webservice to determine if the source is consistent in time and position with a known asteroid. If so, the source is classified as  class {\tt Rock} with high confidence and all other confidences are set to zero. If there is non-negligible proper motion (typically $>0.1$ arcsec/hour) and the ecliptic latitude is small ($b < 15$ deg), then the source is classified as likely class {\tt Rock} (ad hoc, we ascribe a $90$\% confidence to this).
  \item {\it SIMBAD:} About 8.6\% of PTF-discovered sources cross-match with SIMBAD. Some of those types\footnote{See \url{http://simbad.u-strasbg.fr/simbad/sim-display?data=otypes}.} are definitive statements about the class of variability (such as ``EB*'' for eclipsing binary star, ``Mira,'' ``BLLac,'' and ``YSO''). Other SIMBAD types are useful in determining whether the source is galactic or extragalactic in nature (``GinGroup'' for galaxy in group, ``V*'' for variable star). Some SIMBAD ``types'' are ambiguous (e.g., ``Pec*'' for peculiar star, ``Radio,'' and ``Blue''). For a source near (but not consistent with the center of) a SIMBAD-designated galaxy, we label the source {\tt SN/Nova}. 
   \item {\it SDSS:} Spectroscopic redshifts (found in {\tt best\_z}) and galaxy/star separation (based on the PSF of the host) were used as reliable sources of the extragalactic/galactic nature of the PTF source. The spectral typing ({\tt sdss\_spectral\_stellar\_type}) we use to determine the nature of extragalactic events (i.e., labels as {\tt QSO} were taken as definitive).  Hosts labelled as ``star'' but with X-ray or radio matches ({\tt rosat\_cps} and {\tt first\_flux\_in\_mJy}) were taken as likely QSOs.
   \item {\it USNO-B1.0:} host color, offset, and star/galaxy classification are used to make decisions about the extragalactic or galactic nature of the source. Astrometric coincidence with the centers of putative host galaxies are labelled as  {\tt AGN-cnSN-TDE}.
\end{enumerate}
We used a hand-tuned aggregate weighting of all available authorities to produce a single set of confidence statements about the nature of the variables. Internal Oarical discoveries with high real-bogus ($>0.3$) and high classification confidence are saved automatically through the web interface of PTF Marshal, thus assigning an official PTF name and an initial type to the source. When more refined classifications are available (e.g., from SIMBAD or SDSS spectroscopy)  that class is also annotated to the PTF databases as a value-added classification (Figure \ref{fig:taxonomy}).

\begin{figure*}[tbp]
	\centerline{\includegraphics[width=5.5in,angle=0]{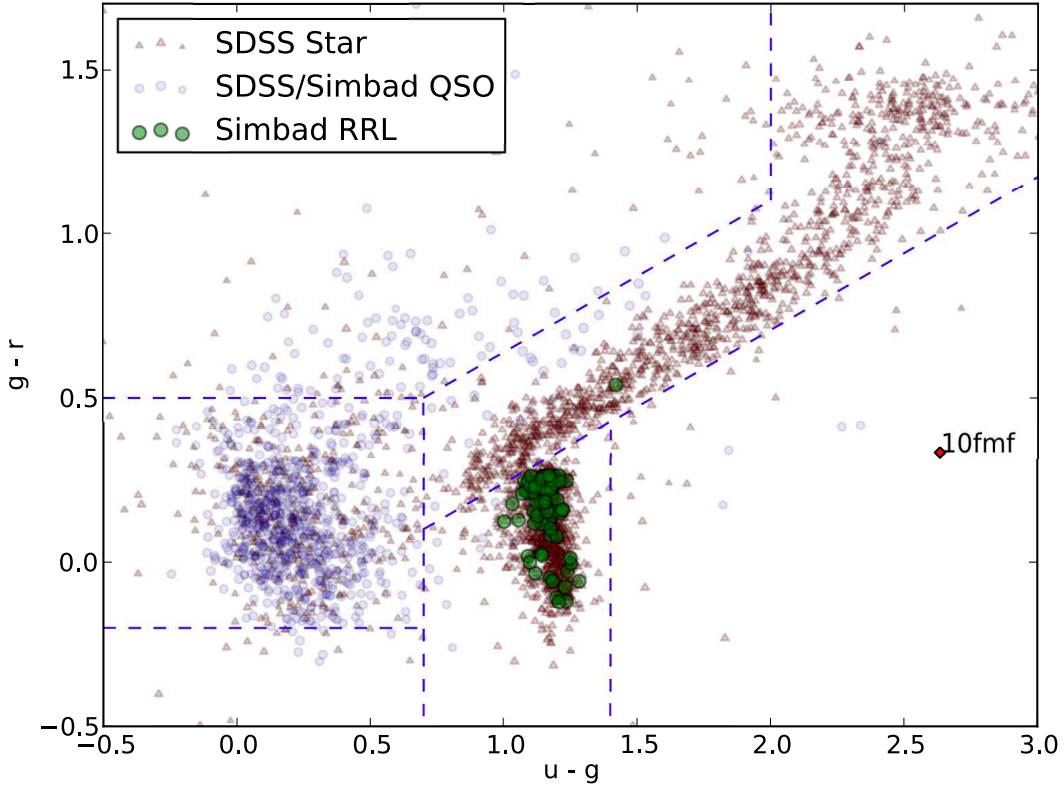}}
\caption[rb]{SDSS color-color diagram of the hosts (labelled in the SDSS photometric table as ``star'') of 3979 PTF sources observed until June 2010. Oarical was used to type and classify (Figure \ref{fig:taxonomy}) these sources using SDSS and SIMBAD. The dashed lines show the regions traditionally used to classify sources (see \citealt{2010ApJ...708..717S}). Known QSOs are shown in blue circles. Cataloged ``stars'' are shown in red triangles. Most of the ``stars'' in the QSOs locus are likely quasars without spectroscopy. Known RR Lyrae stars from SIMBAD are shown with green circles. PTF 10fmf is discussed in the text.}
\label{fig:sdss}
\end{figure*}

Figure \ref{fig:sdss} shows the subset of the Oarical-classified PTF sources with a putative quiescent counterpart in SDSS. The stellar-, QSO-, and RR Lyrae-loci are seen and the density of variables is qualitatively similar to that seen in the Stripe 82 survey of variable sources \citep{2003MmSAI..74..978I}---that is, the relatively rare blue sources tend to be more significantly variable than red stars. There are 78 known RR Lyrae in this sample (from SIMBAD) with an additional 1502 sources matching the color locus of RR Lyrae suggested in \citet{2010ApJ...708..717S}---of these, there are 8 known QSOs matching the RR Lyrae colors\footnote{Two of these are misclassified spectroscopically and are indeed likely RR Lyrae.}.
Since the locus of high-redshift quasars cuts across the RR Lyrae locus \citep{2010ApJ...708..717S} to larger $u - g$ color at roughly constant $g - r$, we decided to obtain a spectrum of one variable ``star'' (PTF 10fmf = SDSS J173630.59+642308.5; $u-g = 2.6$ and $g - r = 0.33$) to ascertain whether PTF was indeed discovering high-redshift QSOs redward of the RRL locus. The spectrum taken with the Keck1+LRIS on June 14, 2010 UT of PTF 10fmf revealed a broad-line QSO at $z=3.2$, making this source the highest redshift PTF-discovered transient.

At the time of writing, there are roughly 40,000 sources discovered by Oarical and stored in the internal Oarical databases. There a total of 28,078 sources in the PTF Marshal database, with 20,355 discoveries or rediscoveries\footnote{A rediscovery is when a scanner (human or robotic) saves a candidate into the PTF Marshal which is associated with a source already previously saved/discovered by the PTF collaboration.} since Oarical began running autonomously. Oarical accounts for 14,466 automatic discoveries or rediscoveries---that is, 29\% of PTF sources were only discovered by human scanners while Oarical was running and about 36\% of Oarical internal discoveries are saved automatically in the PTF Marshal without any humans in the loop.  The other two-thirds of those Oarical sources that are not high-enough quality in score to warrant an automated discovery are presented to human scanners who decide on whether possible new events should be promoted to ``discovery'' (see \S \ref{sec:query} and Figure \ref{fig:oarweb}). We do not currently record when a human discovery was assisted directly by an Oarical-generated webpage---the majority of the 29\% of human-scanner discovery likely originates from the Oarical-generated webpages of possible candidates. The SupernovaZoo accounts for the human-generated discovery of many nearby SNe \citep{2011MNRAS.412.1309S}.

Of the Oarical-discovered sources of PTF type {\sc VarStar}, there were 79 spectroscopic observations recorded in the PTF Marshal  (usually obtained after the Oarical discovery). Twenty two (28\%) of these were spectroscopically typed to be supernovae---that is, incorrectly typed by Oarical. Interestingly, 14 of these had SDSS host identifications as ``star'' and almost all hosts appeared to be very large galaxies where the SDSS source classification broke up the large host into smaller subregions classified incorrectly as stars\footnote{Improved sky-subtraction in SDSS may alleviate some of this problem in the future \citep{2011arXiv1105.1960B}.}.

Of the Oarical discovered sources of PTF type {\sc Transient}, there were 645 sources with spectroscopic observations recorded in the PTF Marshal. Of these, there were 529 sources spectroscopically classified as supernovae, 43 were classified as variable stars, 23 identified as cataclysmic variables, 37 as some type of AGN, and the remaining 39 as unknown or uncertain (26 had more than one classification recorded that differed between categories). That is, about 7\% of Oarical-discovered {\sc Transients} were definitely incorrect.  Since Oarical began, there were a total of 740 sources spectroscopically classified as a supernova of which 535 (72\%) were discovered or rediscovered by Oarical. 

\subsection{Query Mechanisms}
\label{sec:query}

\begin{figure*}[p]
	\centerline{\includegraphics[width=6in,angle=0]{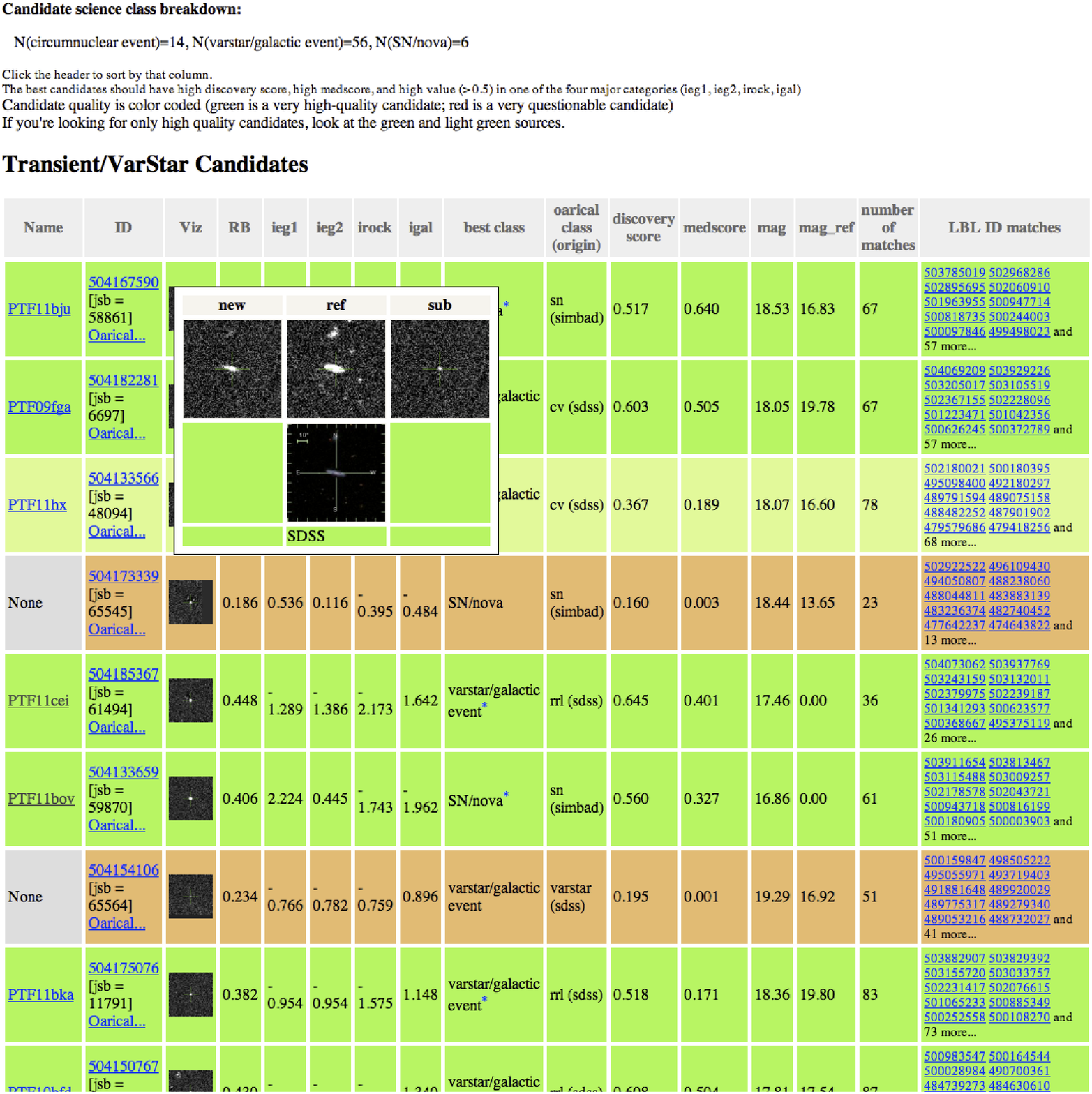}}
\caption[rb]{Screenshot of a webpage generated for human-scanners to view possible candidates for discovery. Previously discovered sources are named following the PTF naming convention, all others are labelled as ``None.'' (on this page there are two previously unknown sources). Color coding of each row shows the relative confidence in the source as a true astrophysical event. The sources with blue asterisks (in ``best class'' column) are ones that Oarical has discovered. When the user mouses over the image thumbnail, a pop up of the subtraction is shown. About 20--30 such pages are generated nightly. Oarical-assisted, human discoveries originate from  these pages.}
\label{fig:oarweb}
\end{figure*}

For all sources saved in the PTF Marshal databases, whether or not Oarical discovered the source, Oarical is run as an annotation service in near real-time. Information from SIMBAD and SDSS are saved as Comments in the PTF Marshal (Fig.\ \ref{fig:oarweb}) and available for users interested a particular source to get a detailed set of metrics (if available) about that place in the sky. For instance, if a user saves a source as a {\sc VarStar} but SDSS has a spectrum of that source, within about 15 minutes, the Marshal will have annotations related to the SDSS spectrum (e.g., whether it is a quasar, what the spectroscopic redshift is determined to be, what the errors on redshift is, etc.). At the time that the source is annotated, any positional coincidence with known IAU Circular supernovae is also marked up in the PTF Marshal.

In addition to the automatic discovery of sources, Oarical provides webpage summaries of possible new sources from each reduction run. An email is sent to PTF subscribers about 30 min after the data are obtained allowing quick perusal of possible new sources; this allows humans to save sources which might not otherwise meet the thresholds for automatic discovery (see fig.\ \ref{fig:oarweb}). A duty astronomer (primarily at the Weizmann Institute of Science) manually scans the Oarical discoveries and possible discoveries every day, in near real-time and assigns followup priorities \citep{ptf10vdl}. 

A webbased interface to Oarical is available to the PTF collaboration. This allows a PTF source, position, or candidate ID to be analyzed even if Oarical has not ingested that source into the databases.  In addition, Oarical is automatically queried about once an hour for recently active sources that meet the criteria of certain science key projects. Fast transients (for example, changing by more than 0.5 magnitude in less than 3 hours) and tidal disruption candidates (circumnuclear events atop quiescent galaxies) have custom webviews autogenerated during the night based on these queries.

\subsection{Machine-Learned Classification}
\label{sec:ml}

\begin{figure*}[tbp]
	\centerline{\includegraphics[width=4.5in,angle=0]{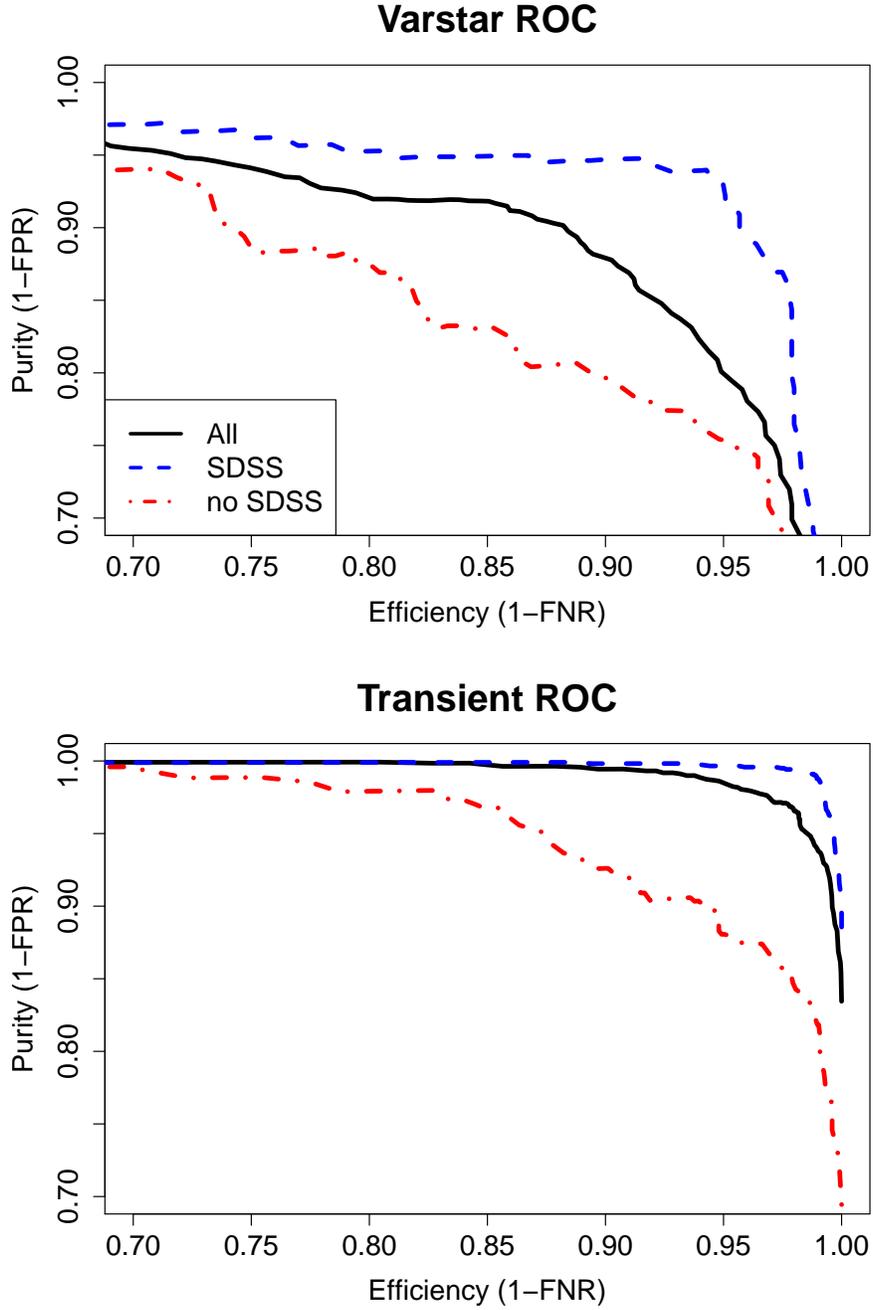}}
\caption[rb]{ROC curves for PTF Type classification.  For each of variable star (top) and transient (bottom) classification, we plot the efficiency and purity of the random forest classifier as a function of the probability threshold.  For the sample of objects used, we recover $\sim$80\% of variable stars and $\sim$99\% of the transient sources at a purity level of 90\%.  The ROC curves for SDSS objects (blue dashed) dominate those for non-SDSS objects (red dot-dash).
}
\label{roc_ptfClass_discovery}
\end{figure*}

With an eye to eventually replacing the manually tuned classification algorithm, we have explored the feasibility of using machine-learned classification for immediate PTF source classification.  Using a sample of 1953 PTF sources with either spectroscopically-confirmed or SIMBAD-determined class, we train a random forest classifier  \citep{2001brei} to predict class as a function of 43 different features.  These features include 9 derived from the PTF light curves and 35 context features.  The random forest classifier operates by constructing an ensemble of classification decision trees, and subsequently averaging the result.  The key to the good performance of random forest is that its component trees are \emph{de-correlated} by sub-selecting a small random number of features as splitting candidates in each non-terminal node of the tree.  As a result, the average of the de-correlated trees has highly decreased variance over each single tree. To handle missing feature values---which arise due to incompleteness in the context features---we use the {\tt missForest} imputation method of \citet{2011arXiv1105.0828S}, which estimates the value of each missing feature via an iterative nonparametric approach to minimize imputation error.
\begin{figure*}[tbp]
	\centerline{\includegraphics[width=4.5in,angle=0]{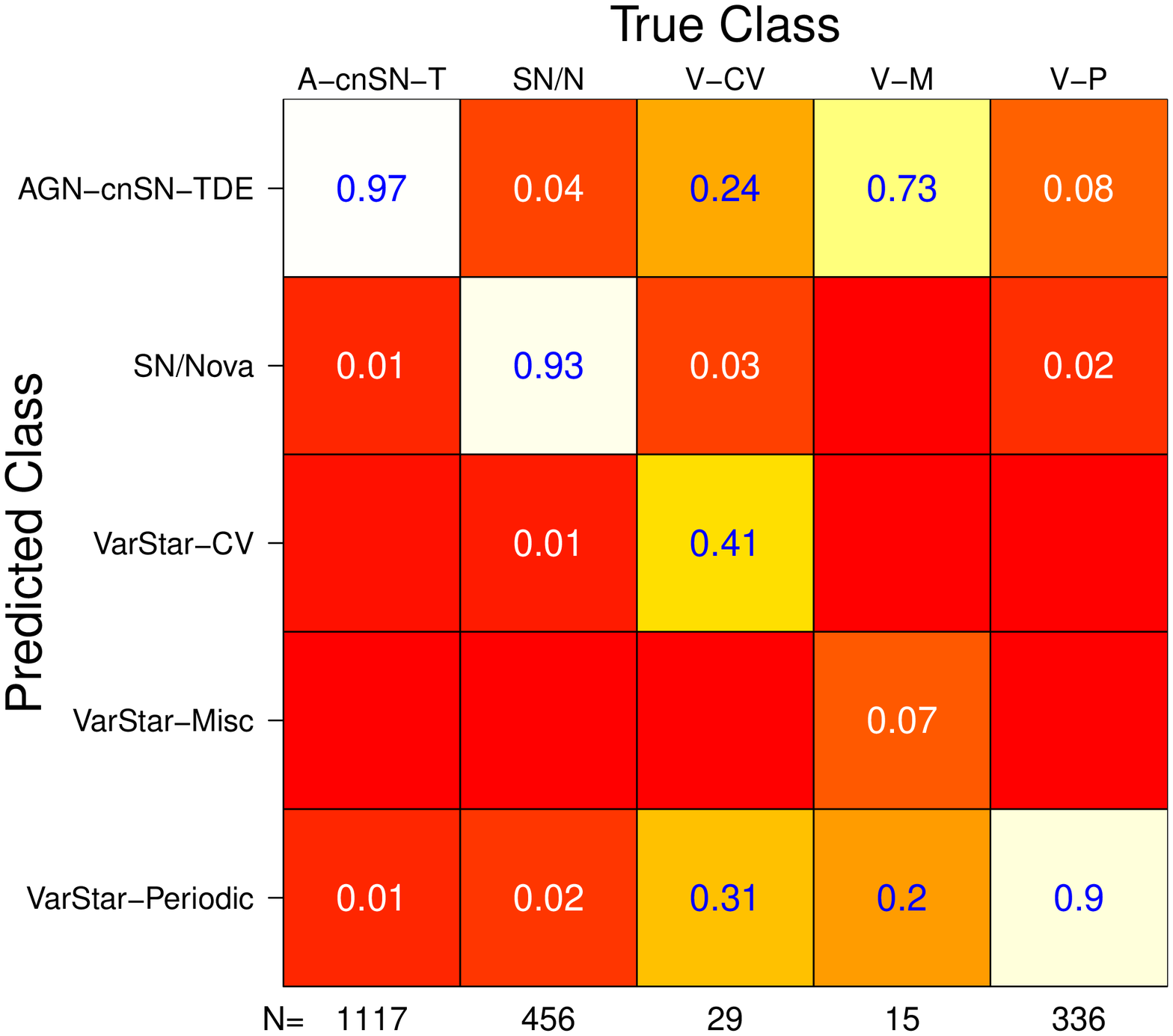}}
\caption[rb]{Confusion matrix for robotclass random forest classification.  Classes are aligned so that entries along the diagonal corresponds to correct classification.  Probabilities are normalized to sum to unity for each column.  Recovery rates are $\ge$90\%, with very high purity, for the three dominant classes.  Classification accuracy suffers for the two classes with small amounts of data (note: class size is written along the bottom of the figure).
}
\label{confmat_robotClass_discovery}
\end{figure*}

For the PTF Type classification problem, we have 1573 {\sc Transient} and 380 {\sc VarStar} sources\footnote{There is some ambiguity in the initial typing scheme in the boundary between {\sc VarStar} and {\sc Transient}: cataclysmic variables (CVs), for instance, could be considered in either category. However, for definiteness, we put CVs in the {\sc VarStar} category.}.  Using features derived at the time of discovery, we obtain a 3.8\% overall error rate (all error rates stated are found using 10-fold cross validation).  For the 1422 sources with SDSS coverage, the error rate is 1.7\%, while for the other 531 sources with no SDSS coverage the error rate jumps to 9.4\%.  In Figure \ref{roc_ptfClass_discovery} we plot the ROC curves for both variable star and transient source classification.  The ROC curves show that at 90\% purity, the random forest classifier attains 96.6\% efficiency of variable star classification and 99.7\% efficiency of transient classification.  Notably, for SDSS sources, we achieve a 96.6\% (100\%) efficiency of {\sc VarStar} ({\sc Transient}) classification at a 90\% purity level.

In robotclass classification, which divides the sources into five science classes, random forest obtains an error rate of 6.5\%.  Figure \ref{confmat_robotClass_discovery} shows that for the {\tt AGN-cnSN-TDE}, {\tt SN/Nova}, and {\tt VarStar-Periodic} classes the classifier attains 97\%, 93\%, and 89\% recovery, respectively.  Due to a large class imbalance, performance of the classifier suffers for the smaller classes {\tt VarStar-CV} and {\tt VarStar-misc}.    Again, our classifier performs significantly better for sources in SDSS, attaining a 3.7\% error rate compared to 14.1\% error for sources with no SDSS coverage.  As more data are collected (post time of discovery), the robotclass random forest error rate decreases slightly: the error rate for objects without SDSS coverage drops to 13.2\% after 30 days and 12.8\% after 90 days, while the error rate for objects in SDSS does not change significantly with increased PTF observations.  This implies that additional PTF data only helps in classification when no SDSS features are available.

\begin{figure*}[tbp]
	\centerline{\includegraphics[width=6.0in,angle=0]{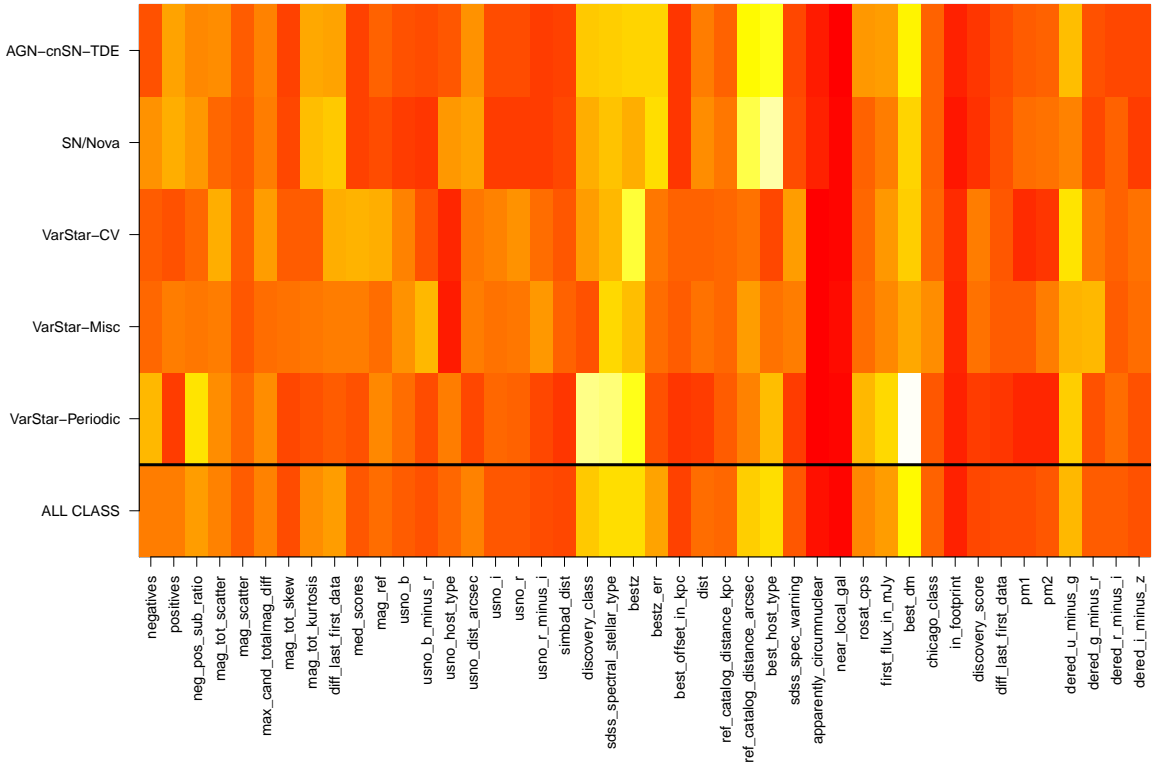}}
\caption[rb]{	Importance of each feature, determined using a pair-wise decision tree algorithm, for classifying objects of each class.  Importance ranges from red (low) to yellow (high).  The average importance across classes shows that PTF light curve features have high importance in the classifier.
}
\label{featImp_robotClass_discovery}
\end{figure*}

Finally, the RF classification trees allow us to construct an estimate of the importance of each feature in the classifier.  Using the prescription of \citet{2001brei}, we compute the importance of each feature as the increased number of sources that are correctly classified when using that feature instead of a replacement feature of random noise.  In Figure \ref{featImp_robotClass_discovery} we plot the importance of each feature for each of the robotclass classes ({\tt VarStar-misc} was omitted due to a scarce amount of data), and the average importance across all classes. Overall, the most important features are context based, while some light-curve-derived features (such as the ratio of the number of negative subtractions to positive subtractions) are important for distinguishing between certain classes. In the future, we may add more descriptive time-series features (such as those related to periodograms) which should also be useful in classification.

There are some biases in the sample generation that require a careful interpretation of these ML results. For a source to be included in the training sample via existing catalogs, it must have a SIMBAD label (e.g., ``RRLyr*'' or ``QSO'') that provides a ``definitive'' ground-truth statement about the nature of the variability. In some cases, that SIMBAD label comes from SDSS spectroscopy (particularly for quasars); since SDSS spectroscopy is used in the ML classification, the information in some of the training set is essentially known perfectly in the classifier (this is one explanation why classification is inferior in non-SDSS footprint fields). Also, SIMBAD sources tend to be brighter than many PTF sources and so the above analysis can be thought of as applying to the brighter end of the distribution. Spectroscopically confirmed SNe candidates found in PTF which are used in the training are obtained after humans in the PTF collaboration have vetted the PTF image-difference-based discoveries and decided to pursue spectroscopic followup. A bright supernova that Oarical (or humans) initially type as {\tt VarStar} might not be inspected by humans and therefore not receive a spectroscopic classification. Likewise, if a source is initially labelled as an SN but a human decides not to pursue spectroscopic followup because the candidate is of poor or dubious quality then that source will not be included in the ML training sample. In this sense, the ML results should (conservatively) be viewed as classification results {\it given that the source} is a) observed to vary significantly in the image differences and b) is a bona fide astrophysical variable or transient.

\section{Discussion and Conclusions}
\label{sec:disandconc}

We have described the framework for building discovery and classification on astronomical synoptic survey streams without humans in the real-time loop. Some features of this framework have been employed previously but, to the best of our knowledge, this is the first example of such an end-to-end framework working in (near) real-time and with real-world data. The use of Oarical in PTF is part an even more expansive thrust of the project in that:
\begin{enumerate}
  \item the data themselves are acquired on an autonomously operated telescope with a computer-generated observing schedule \citep{2009PASP..121.1395L};
  \item images are transported, reduced, and photometered in near-real time \citep{2009PASP..121.1395L};
  \item discovery and classification results are marked up in a central PTF-wide database;
  \item triggers are then generated for followup by autonomous robotic telescopes (namely P60 and PAIRITEL), which followup some high-priority {\sc Transient} sources without humans in the loop \citep{2011arXiv1103.0779C,ptf10vdl}.
\end{enumerate}
\noindent There is, in this sense, a recognition that follow-up of time-variable sources is crucial for the scientific impact in many domains of interest to the PTF community. Autonomous discovery and classification allows for the initial imaging follow-up to be conducted without astronomers in the real-time loop. Our collaboration also routinely conducts (human-intensive) spectroscopic followup on newly discovered Oarical sources with minimal turnaround times from PTF image to spectroscopy to inference. For instance, we obtained with Keck a spectrum on a newly discovered {\sc Transient} 29 minutes after Oarical discovery. The source was a peculiar Type Ia supernova at a redshift $z=0.18,$ and analysis of the spectrum was published less than 18 hours after it was first observed with PTF \citep{atel2600}. \citet{ptf10vdl} gives a full description of rapid discovery, follow-up, and 
the scientific results with PTF.

The 529 spectroscopically-confirmed SNe discovered autonomously by Oarical since April 2010 represent more than half of the SNe discovered by the PTF collaboration over the lifetime of the project. Several key papers have been the result of Oarical discoveries, including discoveries and real-time classification of a) PTF 10iya, a possible tidal disruption event \citep{2011arXiv1103.0779C}, b) PTF 10vdl, a subluminous type IIP supernova \citep{ptf10vdl}, c) PTF 10qpf, a TTauri star that appeared to be an FUOri system in outburst \citep{2011ApJ...730...80M}, d)  PTF 10nvg, an outbursting Class I protostar \citep{2011AJ....141...40C}, and e) PTF 10hmv, a type Ia supernova found more than 10 days before maximum and observed with the Hubble Space telescope around maximum light \citep{2011ApJ...727L..35C}.

The core discovery and classification codebase has been largely frozen since April 2010 allowing us to study the results under the assumption of relative uniformity. However there are several aspects of the framework that we have identified where improvements could be made in future versions (with PTF or otherwise). First, we now have a good deal more ground-truth events in the PTF database that we know are real astrophysical candidates. This larger training set, coupled with new shaped based metrics on the image differences, should much improve the Type I and Type II errors on the discovery front \citep{sahand}. Second, there has been much improvement in the astrometric tie of PTF to SDSS (as well as an expanding footprint of public SDSS imaging), which should continue to improve the reliability of distance-to-host features. Third, the database-based photometry used to calculate the time-series features is known to be suboptimal. New routines  developed within the collaboration can now allow automated forced-aperture and PSF photometry at the candidate positions. Last, we have now approached a regime where there are enough known classes of sources (from SNe to variable star types) that reliable cross-validated classification can be employed to run machine-learned classifications instead of the manually tuned classification algorithm (\S \ref{sec:class}). It is clear from \S \ref{sec:ml} that ML-based classifications are reasonably predictive, with {\sc Transient}/{\sc VarStar} classification errors at the 5\% level. 

It is clearly early days for large-scale discovery and classification frameworks for synoptic astronomical surveys. As we look to future implementations, there are several avenues and questions to explore:
\begin{itemize}
  \item How do we efficiently discover and classify anomalous sources, those that do not easily fit into the classification categories? Likewise, how can we implement something like a matched-filter discovery of certain classes of sources that have predicted optical light curves but have not been observed before? 
  \item What should be the unique roles for citizen scientists in the real-time discovery and classification loop; can some forms of citizen-science markups be adequately reproduced by machine-learned codes?
  \item Is there a path to using context information immediately with new surveys without having to train with real-world data? That is, is a full  prior 3D model of the transient and variable universe needed to train a classifier on the expected contextual data of a survey just coming online?
  \item How applicable is the framework detailed herein to other surveys (with different depths, cadences, etc.)? That is, are real-time classification algorithms and codebases more tuned to the PTF survey specifics and idiosyncrasies than we believe?
  \item How can we use PTF-tuned classification models to predict classes of sources discovered in other surveys?  That is, is there a formal ML-based workflow to bootstrap learning into new survey data? Active (expert) learning might be an appropriate path for exploration \citep{2011arXiv1106.2832R}.
  \item Can classification statements be improved markedly as follow-up results are automatically flowed back into a central repository of photometry? We currently do not rerun classification on sources after new data is obtained by the survey.
  \item What mechanisms can we use to build up a feedback loop into the classification models? If a source is labelled a {\tt SN/Nova} but is spectroscopically identified as an RR Lyrae star, how do we automatically learn from our classification mistakes?
  \item When, in the course of a survey, is it appropriate to relearn classification based on previous results from the survey? How can the discovery and classification biases from previous incarnations of the framework be controlled in new learning iterations while maintaining control of systematics that are crucial for determining event rates?  
\end{itemize}
\noindent These are questions and areas of study we expect to explore in the coming years. With each iteration of the framework, we can hope to produce a more complete and robust framework for use in new surveys. We expect that automatic discovery workflows will need to be highly tuned for each survey but ``classification as a service'' should evolve as a more general framework that could be hosted and maintained by third parties. This appears to be the direction that the LSST collaboration is heading.

\acknowledgements

The authors acknowledge the generous support of a CDI grant (\#0941742) from the National Science Foundation. J.S.B. and D.L.S. also thank the Las Cumbres Observatory for support during the early stages of this work. S.B.C.~wishes to acknowledge generous support from Gary and Cynthia Bengier, the Richard and Rhoda Goldman Fund, National Aeronautics and Space Administration (NASA)/Swift\ grant NNX10AI21G, NASA/{\it Fermi} grant NNX1OA057G, and National Science Foundation (NSF) grant AST--0908886. The National Energy Research Scientific Computing
Center, which is supported by the Office of Science
of the U.S.\ Department of Energy under Contract \#DE-AC02-05CH11231, provided staff, computational resources, and data storage for this project.

Some of the data presented herein were obtained at the
W. M. Keck Observatory, which is operated as a scientific partnership among the California Institute of Technology, the University of California, and the National
Aeronautics and Space Administration. The Observatory was made possible by the generous financial support
of the W. M. Keck Foundation. The authors wish to recognize and acknowledge the very significant cultural role
and reverence that the summit of Mauna Kea has always
had within the indigenous Hawaiian community. We are
most fortunate to have the opportunity to conduct observations from this mountain. This research has made use of the VizieR catalogue access tool, CDS, Strasbourg, France \citep{2000A&AS..143...23O}.

{\it Observatories and Facilities}: Palomar: 48 inch (PTF) Palomar: 200 inch (DBSP) Keck (LRIS) 
Apache Point Observatory (2MASS)


\begin{thebibliography}{55}
\expandafter\ifx\csname natexlab\endcsname\relax\def\natexlab#1{#1}\fi

\bibitem[{{Akerlof} {et~al.}(2003){Akerlof}, {Kehoe}, {McKay}, {Rykoff},
  {Smith}, {Casperson}, {McGowan}, {Vestrand}, {Wozniak}, {Wren}, {Ashley},
  {Phillips}, {Marshall}, {Epps}, \& {Schier}}]{2003PASP..115..132A}
{Akerlof}, C.~W. {\it et al.} 2003, \pasp, 115, 132

\bibitem[{{Bailey} {et~al.}(2007){Bailey}, {Aragon}, {Romano}, {Thomas},
  {Weaver}, \& {Wong}}]{2007ApJ...665.1246B}
{Bailey}, S., {Aragon}, C., {Romano}, R., {Thomas}, R.~C., {Weaver}, B.~A., and
  {Wong}, D. 2007, \apj, 665, 1246

\bibitem[{{Becker} {et~al.}(2005){Becker}, {Axelrod}, {Ivezic}, {Lupton},
  {Silvestri}, \& {Rest}}]{2005AAS...207.2629B}
{Becker}, A., {Axelrod}, T., {Ivezic}, Z., {Lupton}, R., {Silvestri}, N., and
  {Rest}, A. 2005, in Bulletin of the American Astronomical Society, Vol.~37,
  American Astronomical Society Meeting Abstracts, 1206

\bibitem[{{Belokurov} {et~al.}(2003){Belokurov}, {Evans}, \&
  {Du}}]{2003MNRAS.341.1373B}
{Belokurov}, V., {Evans}, N.~W., and {Du}, Y.~L. 2003, \mnras, 341, 1373

\bibitem[{{Bertin} \& {Arnouts}(1996)}]{1996A&AS..117..393B}
{Bertin}, E. and {Arnouts}, S. 1996, \aaps, 117, 393

\bibitem[{{Blanton} {et~al.}(2011){Blanton}, {Kazin}, {Muna}, {Weaver}, \&
  {Price-Whelan}}]{2011arXiv1105.1960B}
{Blanton}, M.~R., {Kazin}, E., {Muna}, D., {Weaver}, B.~A., and {Price-Whelan},
  A. 2011, {arXiv/1105.1960}

\bibitem[{{Bloom} \& {Richards}(2011)}]{2011arXiv1104.3142B}
{Bloom}, J.~S. and {Richards}, J.~W. 2011, {arXiv/1104.3142}

\bibitem[{{Bloom} {et~al.}(2006){Bloom}, {Starr}, {Blake}, {Skrutskie}, \&
  {Falco}}]{2006ASPC..351..751B}
{Bloom}, J.~S., {Starr}, D.~L., {Blake}, C.~H., {Skrutskie}, M.~F., and
  {Falco}, E.~E. 2006, in Astronomical Society of the Pacific Conference
  Series, Vol. 351, Astronomical Data Analysis Software and Systems XV, ed.
  {C.~Gabriel, C.~Arviset, D.~Ponz, \& S.~Enrique}, 751--+

\bibitem[{{Bond} {et~al.}(2001){Bond}, {Abe}, {Dodd}, {Hearnshaw}, {Honda},
  {Jugaku}, {Kilmartin}, {Marles}, {Masuda}, {Matsubara}, {Muraki}, {Nakamura},
  {Nankivell}, {Noda}, {Noguchi}, {Ohnishi}, {Rattenbury}, {Reid}, {Saito},
  {Sato}, {Sekiguchi}, {Skuljan}, {Sullivan}, {Sumi}, {Takeuti}, {Watase},
  {Wilkinson}, {Yamada}, {Yanagisawa}, \& {Yock}}]{2001MNRAS.327..868B}
{Bond}, I.~A. {\it et al.} 2001, \mnras, 327, 868

\bibitem[{Breiman(2001)}]{2001brei}
Breiman, L. 2001, Machine learning, 45, 5

\bibitem[{{Butler} \& {Bloom}(2010)}]{2010arXiv1008.3143B}
{Butler}, N.~R. and {Bloom}, J.~S. 2010, {arXiv/1008.3143}

\bibitem[{{Castro-Tirado}(2010)}]{ca2010}
{Castro-Tirado}, A.~J. 2010, Advances in Astronomy, 2010

\bibitem[{{Cenko} {et~al.}(2011){Cenko}, {Bloom}, {Kulkarni}, {Strubbe},
  {Miller}, {Butler}, {Quimby}, {Gal-Yam}, {Ofek}, {Quataert}, {Bildsten},
  {Poznanski}, {Perley}, {Morgan}, {Filippenko}, {Arcavi}, {Ben-Ami},
  {Cucchiara}, {Fassnacht}, {Green}, {Hook}, {Howell}, {Lagattuta}, {Law},
  {Kasliwal}, {Nugent}, {Silverman}, {Sullivan}, {Tendulkar}, \&
  {Yaron}}]{2011arXiv1103.0779C}
{Cenko}, S.~B. {\it et al.} 2011, arXiv/1103.0779

\bibitem[{{Cenko} {et~al.}(2006){Cenko}, {Fox}, {Moon}, {Harrison}, {Kulkarni},
  {Henning}, {Guzman}, {Bonati}, {Smith}, {Thicksten}, {Doyle}, {Petrie},
  {Gal-Yam}, {Soderberg}, {Anagnostou}, \& {Laity}}]{2006PASP..118.1396C}
---. 2006, \pasp, 118, 1396

\bibitem[{Ciurana(2009)}]{1507544}
Ciurana, E. 2009, Developing with Google App Engine (Berkeley, CA, USA: Apress)

\bibitem[{{Cooke} {et~al.}(2011){Cooke}, {Ellis}, {Sullivan}, {Nugent},
  {Howell}, {Gal-Yam}, {Lidman}, {Bloom}, {Cenko}, {Kasliwal}, {Kulkarni},
  {Law}, {Ofek}, \& {Quimby}}]{2011ApJ...727L..35C}
{Cooke}, J. {\it et al.} 2011, \apjl, 727, L35

\bibitem[{{Covey} {et~al.}(2011){Covey}, {Hillenbrand}, {Miller}, {Poznanski},
  {Cenko}, {Silverman}, {Bloom}, {Kasliwal}, {Fischer}, {Rayner}, {Rebull},
  {Butler}, {Filippenko}, {Law}, {Ofek}, {Ag{\"u}eros}, {Dekany}, {Rahmer},
  {Hale}, {Smith}, {Quimby}, {Nugent}, {Jacobsen}, {Zolkower}, {Velur},
  {Walters}, {Henning}, {Bui}, {McKenna}, {Kulkarni}, \&
  {Klein}}]{2011AJ....141...40C}
{Covey}, K.~R. {\it et al.} 2011, \aj, 141, 40

\bibitem[{{Debosscher} {et~al.}(2007){Debosscher}, {Sarro}, {Aerts}, {Cuypers},
  {Vandenbussche}, {Garrido}, \& {Solano}}]{2007A&A...475.1159D}
{Debosscher}, J., {Sarro}, L.~M., {Aerts}, C., {Cuypers}, J., {Vandenbussche},
  B., {Garrido}, R., and {Solano}, E. 2007, \aap, 475, 1159

\bibitem[{{Drake} {et~al.}(2009){Drake}, {Djorgovski}, {Mahabal}, {Beshore},
  {Larson}, {Graham}, {Williams}, {Christensen}, {Catelan}, {Boattini},
  {Gibbs}, {Hill}, \& {Kowalski}}]{2009ApJ...696..870D}
{Drake}, A.~J. {\it et al.} 2009, \apj, 696, 870

\bibitem[{{Filippenko} {et~al.}(2001){Filippenko}, {Li}, {Treffers}, \&
  {Modjaz}}]{2001ASPC..246..121F}
{Filippenko}, A.~V., {Li}, W.~D., {Treffers}, R.~R., and {Modjaz}, M. 2001, in
  Astronomical Society of the Pacific Conference Series, Vol. 246, IAU Colloq.
  183: Small Telescope Astronomy on Global Scales, ed. {B.~Paczynski,
  W.-P.~Chen, \& C.~Lemme}, 121

\bibitem[{{Flaugher}(2005)}]{2005IJMPA..20.3121F}
{Flaugher}, B. 2005, International Journal of Modern Physics A, 20, 3121

\bibitem[{{Gal-Yam} \& {Mazzali}(2011)}]{2011arXiv1103.5165G}
{Gal-Yam}, A. and {Mazzali}, P. 2011, {arXiv/1103.5165}

\bibitem[{Gal-Yam {et~al.}(2011)}]{ptf10vdl}
Gal-Yam, A. {\it et al.} 2011, submitted to ApJL

\bibitem[{Hall {et~al.}(2009)Hall, Frank, Holmes, Pfahringer, Reutemann, \&
  Witten}]{weka}
Hall, M., Frank, E., Holmes, G., Pfahringer, B., Reutemann, P., and Witten,
  I.~H. 2009, SIGKDD Explorations, 11, 10

\bibitem[{{Ivezi{\'c}} {et~al.}(2003){Ivezi{\'c}}, {Lupton}, {Anderson},
  {Eyer}, {Gunn}, {Juri{\'c}}, {Knapp}, {Miknaitis}, {Gunn}, {Rockosi},
  {Schlegel}, {Strauss}, {Stubbs}, \& {Vanden Berk}}]{2003MmSAI..74..978I}
{Ivezi{\'c}}, {\v Z}. {\it et al.} 2003, \memsai, 74, 978

\bibitem[{{Ivezic} {et~al.}(2008){Ivezic}, {Tyson}, {Acosta}, {Allsman},
  {Anderson}, {Andrew}, {Angel}, {Axelrod}, {Barr}, {Becker}, {Becla},
  {Beldica}, {Blandford}, {Bloom}, {Borne}, {Brandt}, {Brown}, {Bullock},
  {Burke}, {Chandrasekharan}, {Chesley}, {Claver}, {Connolly}, {Cook},
  {Cooray}, {Covey}, {Cribbs}, {Cutri}, {Daues}, {Delgado}, {Ferguson},
  {Gawiser}, {Geary}, {Gee}, {Geha}, {Gibson}, {Gilmore}, {Gressler}, {Hogan},
  {Huffer}, {Jacoby}, {Jain}, {Jernigan}, {Jones}, {Juric}, {Kahn}, {Kalirai},
  {Kantor}, {Kessler}, {Kirkby}, {Knox}, {Krabbendam}, {Krughoff}, {Kulkarni},
  {Lambert}, {Levine}, {Liang}, {Lim}, {Lupton}, {Marshall}, {Marshall}, {May},
  {Miller}, {Mills}, {Monet}, {Neill}, {Nordby}, {O'Connor}, {Oliver},
  {Olivier}, {Olsen}, {Owen}, {Peterson}, {Petry}, {Pierfederici},
  {Pietrowicz}, {Pike}, {Pinto}, {Plante}, {Radeka}, {Rasmussen}, {Ridgway},
  {Rosing}, {Saha}, {Schalk}, {Schindler}, {Schneider}, {Schumacher}, {Sebag},
  {Seppala}, {Shipsey}, {Silvestri}, {Smith}, {Smith}, {Strauss}, {Stubbs},
  {Sweeney}, {Szalay}, {Thaler}, {Vanden Berk}, {Walkowicz}, {Warner},
  {Willman}, {Wittman}, {Wolff}, {Wood-Vasey}, {Yoachim}, {Zhan}, \& {for the
  LSST Collaboration}}]{2008arXiv0805.2366I}
{Ivezic}, Z. {\it et al.} 2008, {arxiv/0805.2366}

\bibitem[{Jeffreys(1946)}]{Jeffreys24091946}
Jeffreys, H. 1946, Proceedings of the Royal Society of London. Series A.
  Mathematical and Physical Sciences, 186, 453

\bibitem[{{Juri{\'c}} \& {Ivezi{\'c}}(2011)}]{2011EAS....45..281J}
{Juri{\'c}}, M. and {Ivezi{\'c}}, {\v Z}. 2011, in EAS Publications Series,
  Vol.~45, EAS Publications Series, 281--286

\bibitem[{{Kaiser} {et~al.}(2002){Kaiser}, {Aussel}, {Burke}, {Boesgaard},
  {Chambers}, {Chun}, {Heasley}, {Hodapp}, {Hunt}, {Jedicke}, {Jewitt},
  {Kudritzki}, {Luppino}, {Maberry}, {Magnier}, {Monet}, {Onaka}, {Pickles},
  {Rhoads}, {Simon}, {Szalay}, {Szapudi}, {Tholen}, {Tonry}, {Waterson}, \&
  {Wick}}]{2002SPIE.4836..154K}
{Kaiser}, N. {\it et al.} 2002, in Society of Photo-Optical Instrumentation
  Engineers (SPIE) Conference Series, ed. {J.~A.~Tyson \& S.~Wolff}, Vol. 4836,
  154--164

\bibitem[{{Keller} {et~al.}(2007){Keller}, {Schmidt}, {Bessell}, {Conroy},
  {Francis}, {Granlund}, {Kowald}, {Oates}, {Martin-Jones}, {Preston},
  {Tisserand}, {Vaccarella}, \& {Waterson}}]{2007PASA...24....1K}
{Keller}, S.~C. {\it et al.} 2007, \pasa, 24, 1

\bibitem[{{Kubanek}(2010)}]{2010arXiv1002.0108K}
{Kubanek}, P. 2010, {arXiv/1002.0108}

\bibitem[{{Law} {et~al.}(2010){Law}, {Dekany}, {Rahmer}, {Hale}, {Smith},
  {Quimby}, {Ofek}, {Kasliwal}, {Zolkower}, {Velur}, {Henning}, {Bui},
  {McKenna}, {Nugent}, {Jacobsen}, {Walters}, {Bloom}, {Surace}, {Grillmair},
  {Laher}, {Mattingly}, \& {Kulkarni}}]{2010SPIE.7735E.122L}
{Law}, N.~M. {\it et al.} 2010, in Society of Photo-Optical Instrumentation
  Engineers (SPIE) Conference Series, Vol. 7735, Society of Photo-Optical
  Instrumentation Engineers (SPIE) Conference Series

\bibitem[{{Law} {et~al.}(2009){Law}, {Kulkarni}, {Dekany}, {Ofek}, {Quimby},
  {Nugent}, {Surace}, {Grillmair}, {Bloom}, {Kasliwal}, {Bildsten}, {Brown},
  {Cenko}, {Ciardi}, {Croner}, {Djorgovski}, {van Eyken}, {Filippenko}, {Fox},
  {Gal-Yam}, {Hale}, {Hamam}, {Helou}, {Henning}, {Howell}, {Jacobsen},
  {Laher}, {Mattingly}, {McKenna}, {Pickles}, {Poznanski}, {Rahmer}, {Rau},
  {Rosing}, {Shara}, {Smith}, {Starr}, {Sullivan}, {Velur}, {Walters}, \&
  {Zolkower}}]{2009PASP..121.1395L}
{Law}, N.~M. {\it et al.} 2009, \pasp, 121, 1395

\bibitem[{{Mahabal} {et~al.}(2008){Mahabal}, {Djorgovski}, {Williams}, {Drake},
  {Donalek}, {Graham}, {Moghaddam}, {Turmon}, {Jewell}, {Khosla}, \&
  {Hensley}}]{2008AIPC.1082..287M}
{Mahabal}, A. {\it et al.} 2008, in American Institute of Physics Conference
  Series, Vol. 1082, American Institute of Physics Conference Series, ed.
  {C.~A.~L.~Bailer-Jones}, 287--293

\bibitem[{{Miller} {et~al.}(2011){Miller}, {Hillenbrand}, {Covey}, {Poznanski},
  {Silverman}, {Kleiser}, {Rojas-Ayala}, {Muirhead}, {Cenko}, {Bloom},
  {Kasliwal}, {Filippenko}, {Law}, {Ofek}, {Dekany}, {Rahmer}, {Hale}, {Smith},
  {Quimby}, {Nugent}, {Jacobsen}, {Zolkower}, {Velur}, {Walters}, {Henning},
  {Bui}, {McKenna}, {Kulkarni}, {Klein}, {Kandrashoff}, \&
  {Morton}}]{2011ApJ...730...80M}
{Miller}, A.~A. {\it et al.} 2011, \apj, 730, 80

\bibitem[{{Miller} {et~al.}(2010){Miller}, {Smith}, {Li}, {Bloom}, {Chornock},
  {Filippenko}, \& {Prochaska}}]{2010AJ....139.2218M}
{Miller}, A.~A., {Smith}, N., {Li}, W., {Bloom}, J.~S., {Chornock}, R.,
  {Filippenko}, A.~V., and {Prochaska}, J.~X. 2010, \aj, 139, 2218

\bibitem[{Negahban {et~al.}(2011)Negahban, Poznanski, {et~al.}}]{sahand}
Negahban, S., Poznanski, D., {\it et al.} 2011, in preparation

\bibitem[{{Nugent} {et~al.}(2010){Nugent}, {Cenko}, {Miller}, {Poznanski},
  {Bloom}, {Filippenko}, {Sullivan}, {Howell}, {Quimby}, {Ofek}, {Kasliwal},
  {Kulkarni}, {Law}, {Dekany}, {Rahmer}, {Hale}, {Smith}, {Zolkower}, {Velur},
  {Walters}, {Henning}, {Bui}, {McKenna}, \& {Jacobsen}}]{atel2600}
{Nugent}, P. {\it et al.} 2010, The Astronomer's Telegram, 2600, 1

\bibitem[{{Ochsenbein} {et~al.}(2000){Ochsenbein}, {Bauer}, \&
  {Marcout}}]{2000A&AS..143...23O}
{Ochsenbein}, F., {Bauer}, P., and {Marcout}, J. 2000, \aaps, 143, 23

\bibitem[{{Rau} {et~al.}(2009){Rau}, {Kulkarni}, {Law}, {Bloom}, {Ciardi},
  {Djorgovski}, {Fox}, {Gal-Yam}, {Grillmair}, {Kasliwal}, {Nugent}, {Ofek},
  {Quimby}, {Reach}, {Shara}, {Bildsten}, {Cenko}, {Drake}, {Filippenko},
  {Helfand}, {Helou}, {Howell}, {Poznanski}, \&
  {Sullivan}}]{2009PASP..121.1334R}
{Rau}, A. {\it et al.} 2009, \pasp, 121, 1334

\bibitem[{Richards \& et~al.(2011)}]{2011rich}
Richards, J.~W. and et~al. 2011, {arXiv/1101.1959}

\bibitem[{{Richards} {et~al.}(2011){Richards}, {Starr}, {Brink}, {Miller},
  {Bloom}, {Butler}, {Berian James}, {Long}, \& {Rice}}]{2011arXiv1106.2832R}
{Richards}, J.~W. {\it et al.} 2011, {arxiv/1106.2832}

\bibitem[{{Sarro} {et~al.}(2009){Sarro}, {Debosscher}, {L{\'o}pez}, \&
  {Aerts}}]{2009A&A...494..739S}
{Sarro}, L.~M., {Debosscher}, J., {L{\'o}pez}, M., and {Aerts}, C. 2009, \aap,
  494, 739

\bibitem[{{Sarro} {et~al.}(2006){Sarro}, {S{\'a}nchez-Fern{\'a}ndez}, \&
  {Gim{\'e}nez}}]{2006A&A...446..395S}
{Sarro}, L.~M., {S{\'a}nchez-Fern{\'a}ndez}, C., and {Gim{\'e}nez}, {\'A}.
  2006, \aap, 446, 395

\bibitem[{{Saunders} {et~al.}(2008){Saunders}, {Naylor}, \&
  {Allan}}]{2008AN....329..321S}
{Saunders}, E.~S., {Naylor}, T., and {Allan}, A. 2008, Astronomische
  Nachrichten, 329, 321

\bibitem[{{Sesar} {et~al.}(2010){Sesar}, {Ivezi{\'c}}, {Grammer}, {Morgan},
  {Becker}, {Juri{\'c}}, {De Lee}, {Annis}, {Beers}, {Fan}, {Lupton}, {Gunn},
  {Knapp}, {Jiang}, {Jester}, {Johnston}, \& {Lampeitl}}]{2010ApJ...708..717S}
{Sesar}, B. {\it et al.} 2010, \apj, 708, 717

\bibitem[{{Smith} {et~al.}(2011){Smith}, {Lynn}, {Sullivan}, {Lintott},
  {Nugent}, {Botyanszki}, {Kasliwal}, {Quimby}, {Bamford}, {Fortson},
  {Schawinski}, {Hook}, {Blake}, {Podsiadlowski}, {J{\"o}nsson}, {Gal-Yam},
  {Arcavi}, {Howell}, {Bloom}, {Jacobsen}, {Kulkarni}, {Law}, {Ofek}, \&
  {Walters}}]{2011MNRAS.412.1309S}
{Smith}, A.~M. {\it et al.} 2011, \mnras, 412, 1309

\bibitem[{{Soko{\l}owski} {et~al.}(2010){Soko{\l}owski}, {Ma{\l}ek},
  {Piotrowski}, \& {Wrochna}}]{2010AdAst2010E..54S}
{Soko{\l}owski}, M., {Ma{\l}ek}, K., {Piotrowski}, L.~W., and {Wrochna}, G.
  2010, Advances in Astronomy, 2010

\bibitem[{{Stekhoven} \& {B\"uhlmann}(2011)}]{2011arXiv1105.0828S}
{Stekhoven}, D.~J. and {B\"uhlmann}, P. 2011, {arXiv/1105.0828}

\bibitem[{{Sullivan} {et~al.}(2011){Sullivan}, {Kasliwal}, {Nugent}, {Howell},
  {Thomas}, {Ofek}, {Arcavi}, {Blake}, {Cooke}, {Gal-Yam}, {Hook}, {Mazzali},
  {Podsiadlowski}, {Quimby}, {Bildsten}, {Bloom}, {Cenko}, {Kulkarni}, {Law},
  \& {Poznanski}}]{2011ApJ...732..118S}
{Sullivan}, M. {\it et al.} 2011, \apj, 732, 118

\bibitem[{{Tomaney} \& {Crotts}(1996)}]{1996AJ....112.2872T}
{Tomaney}, A.~B. and {Crotts}, A.~P.~S. 1996, \aj, 112, 2872

\bibitem[{{Vestrand} {et~al.}(2002){Vestrand}, {Borozdin}, {Brumby},
  {Casperson}, {Fenimore}, {Galassi}, {McGowan}, {Perkins}, {Priedhorsky},
  {Starr}, {White}, {Wozniak}, \& {Wren}}]{2002SPIE.4845..126V}
{Vestrand}, W.~T. {\it et al.} 2002, in Presented at the Society of
  Photo-Optical Instrumentation Engineers (SPIE) Conference, Vol. 4845, Society
  of Photo-Optical Instrumentation Engineers (SPIE) Conference Series, ed.
  {R.~I.~Kibrick}, 126--136

\bibitem[{{Willemsen} \& {Eyer}(2007)}]{2007arXiv0712.2898W}
{Willemsen}, P.~G. and {Eyer}, L. 2007, arXiv e-print 0712.2898

\bibitem[{{Wozniak}(2000)}]{2000AcA....50..421W}
{Wozniak}, P.~R. 2000, \actaa, 50, 421

\bibitem[{{Yip} {et~al.}(2004){Yip}, {Connolly}, {Szalay}, {Budav{\'a}ri},
  {SubbaRao}, {Frieman}, {Nichol}, {Hopkins}, {York}, {Okamura}, {Brinkmann},
  {Csabai}, {Thakar}, {Fukugita}, \& {Ivezi{\'c}}}]{2004AJ....128..585Y}
{Yip}, C.~W. {\it et al.} 2004, \aj, 128, 585

\end{thebibliography}

\begin{deluxetable}{ccl} 
\rotate
\tablecolumns{3} 
\tablewidth{8in} 
\tablecaption{Realbogus Features} 
\tablehead{ 
\colhead{Feature Name}    & \colhead{Type} &  \colhead{Description}   }
\startdata 
\verb mag     & numeric & USNO-B1.0 derived magnitude of the candidate on the difference image \\
\verb mag_err & numeric & estimated uncertainty on \verb mag \\
\verb a_image & numeric & semi-major axis of the candidate\tablenotemark{a} \\  
\verb b_image & numeric & semi-minor axis of the candidate\tablenotemark{a} \\  
\verb fwhm    & numeric & full-width at half maximum of the candidate \\
\verb flag    & numeric & numerical representation of the SExtractor extraction flags\tablenotemark{a} \\
\verb mag_ref & numeric & magnitude of the nearest object in the reference image if less than \\
              &          & 5 arcsec from the candidate \\
\verb mag_ref_err & numeric & estimated uncertainty on \verb mag_ref \\
\verb a_ref  & numeric & semi-major axis of the reference source\tablenotemark{a} \\
\verb b_ref  & numeric & semi-minor axis of the reference source\tablenotemark{a} \\
\verb n2sig3 & numeric & number of at least negative 2 $\sigma$ pixels in a 5$\times$5 box centered on the candidate \\
\verb n3sig3 & numeric & number of at least negative 3 $\sigma$ pixels in a 5$\times$5 box centered on the candidate \\
\verb n2sig5 & numeric & number of at least negative 2 $\sigma$ pixels in a 7$\times$7 box centered on the candidate \\
\verb n3sig5 & numeric & number of at least negative 3 $\sigma$ pixels in a 7$\times$7 box centered on the candidate \\
\verb nmask & numeric & number of masked (suspect) pixels within a 5$\times$5 box centered on the candidate \\
\verb flux_ratio & numeric & ratio of the aperture flux of the candidate relative to the aperture flux \\
                 &         & of the reference source \\
\verb ellipticity & numeric & ellipticity of the candidate using \verb a_image ~and \verb b_image \\
\verb ellipticity_ref & numeric & ellipticity of the reference source using \verb a_ref ~and \verb b_ref \\
\verb nn_dist_renorm & numeric & distance in arcseconds from the candidate to reference source \\
\verb magdiff & numeric & when a reference source is found nearby, the difference between the candidate \\
              &         & magnitude and the reference source. Else, the difference between the candidate \\
              &         & magnitude and the limiting magnitude of the image\\
\verb maglim & nominal & True if there is no nearby reference source, False otherwise.\\
\verb sigflux & numeric & significance of the detection, the PSF flux divided by the  \\
              &         & estimated uncertainty in the PSF flux \\
\verb seeing_ratio & numeric & ratio of the FWHM of the seeing on the new image to the FWHM  \\
             &           &  of the seeing on the reference image \\
\verb mag_from_limit & numeric & limiting magnitude minus the candidate magnitude \\
\verb normalized_fwhm & numeric & ratio of the FWHM of the candidate to the seeing in the new image \\
\verb normalized_fwhm_ref & numeric & ratio of the FWHM of the reference source to the seeing in the \\
         &   & reference image \\
\verb good_cand_density & numeric & ratio of the number of candidates in that subtraction to the total \\  
        &    & usable area on that array \\
\verb min_distance_to_edge_in_new & numeric & distance in pixels to the nearest edge of the array on the new image \\
\enddata
\tablenotetext{a}{\citet{1996A&AS..117..393B}}
\label{tab:features}
\end{deluxetable}%

\begin{deluxetable}{cl} 
\rotate
\tablecolumns{2} 
\tablewidth{8in} 
\tablecaption{Time-Domain Features Used for Oarical Classification} 
\tablehead{ 
\colhead{Feature Name} &  \colhead{Description}   }
\startdata 
{\tt negatives}     &  number of candidates found in negative image differences associated with the source \\
 &  That is, the number of epochs where the source was fainter than its reference brightness \\
{\tt positives}     & number of candidates found in the image differences associated with the source \\
{\tt neg\_pos\_sub\_ratio} & ratio of the number of {\tt negatives} to all candidates ({\tt negatives} + {\tt positives})\\
{\tt mag\_scatter} & RMS of the image difference magnitudes of {\tt positive} candidates\\
{\tt mag\_tot\_scatter} & RMS of the total aperture photometry of all candidates \\
{\tt max\_cand\_totalmag\_diff} & maximum of the total-aperture magnitude minus the reference image source magnitude
\\
{\tt diff\_last\_first\_data} &  difference in time (units of days) between the first and the last observation \\
 & associated with the source \\
{\tt pm1} & apparent proper motion (arcsecond/hour) between the first and second epoch \\
 & associated with the source\\
{\tt pm2} & apparent proper motion (arcsecond/hour) between the second and second-to-last \\
 & observation of the source\\
\enddata
\label{tab:tdfeatures}
\end{deluxetable}%

\begin{deluxetable}{ccl} 
\rotate
\tablecolumns{3} 
\tablewidth{8in}
\tablewidth{8in} 
\tablecaption{Context Features Used for Oarical Classification} 
\tablehead{ 
\colhead{Feature Name} &  \colhead{Type} & \colhead{Description}}
\startdata
\hline

\hline
\cutinhead{USNO-B1.0 Based}
{\tt usno\_b} & numeric & B-band magnitude of the nearest source within 5\arcsec\ \\
{\tt usno\_i} & numeric & I-band magnitude of the nearest source within 5\arcsec\ \\
{\tt usno\_r} & numeric & R-band magnitude of the nearest source within 5\arcsec\ \\
{\tt usno\_b\_minus\_r} & numeric & B-band minus R-band magnitude of the nearest \\
   & &  source within 5\arcsec\ \\
{\tt usno\_r\_minus\_i} & numeric & R-band minus I-band magnitude of the nearest \\
   & &  source within 5\arcsec\ \\
{\tt usno\_host\_type} & nominal & Based on the average of the star/galaxy index (``s/g'') \\ & & USNO-B1.0\tablenotemark{a}. Set to ``galaxy'' if s/g $<$ 3.8, ``star'' if \\ 
 & & s/g $>$ 6.7 and, otherwise, ``uncertain'' \\
\hline
\cutinhead{SDSS DR7 Based}
{\tt in\_footprint} & nominal & Position is in the SDSS DR7 footprint (``yes'' or ``no'') \\
{\tt dist\_in\_arcmin} & nominal & distance in arcminutes of the source from the SDSS catalog position\\
{\tt dered\_u\_minus\_g} & numeric & dereddened $u$ minus $g$ magnitude of the nearest source \\
{\tt dered\_g\_minus\_r} & numeric & dereddened $g$ minus $r$ magnitude of the nearest source \\
{\tt dered\_r\_minus\_i} & numeric & dereddened $r$ minus $i$ magnitude of the nearest source \\
{\tt dered\_i\_minus\_z} & numeric & dereddened $i$ minus $z$ magnitude of the nearest source \\
{\tt chicago\_class} & numeric & galaxy principal component classification\tablenotemark{b} \\
{\tt best\_z} & numeric & best redshift available: spectroscopic when {\tt SpecObjAll.zConf} \\
             &          & flag is $>0.5$ \\
            & & {\tt photoz2.photozcc2} when the $r$ magnitude of the reference \\
           &  &  source $>$ 20, \\
            & & {\tt photoz2.photozd1} when the $r$ magnitude of the reference  \\
            & & source $\le$ 20, \\
            & & {\tt photoz.z} otherwise \\
{\tt best\_z\_err} & numeric & uncertainty in the {\tt best\_z} \\
{\tt best\_dm} & numeric & distance modulus (mag) associated with the {\tt best\_z} \\
{\tt best\_offset\_in\_kpc} & numeric & projected physical offset in kpc \\
  & & from {\tt dist\_in\_arcmin} and {\tt best\_z} \\
{\tt first\_flux\_in\_mJy} & numeric & 21cm flux in mJy based on a cross-match with the FIRST survey \\
{\tt rosat\_cps} & numeric &counts per second of the cross-matched source in the ROSAT \\
                 &        &  All-Sky Survey \\
{\tt sdss\_spectral\_stellar\_type} & nominal & spectroscopic classification ({\tt sppParam.sptypea})\tablenotemark{c} \\ 
{\tt sdss\_spec\_warning} & list of nominal & spectroscopic flags related to classification\tablenotemark{d} \\
\cutinhead{PTF and Local Galaxy Catalog Based}
{\tt nn\_dist} & numeric & Distance of the nearest source in the reference image in \\
 & &   arcseconds (if $<10$\arcsec), and unknown otherwise \\
{\tt nn\_kpc} & numeric & Distance of the nearest source in the reference image in \\
 & &   kpc (if {\tt nn\_dist} $<10$\arcsec\ and {\tt bestz} $>$ 0.0001), and unknown  otherwise \\
{\tt near\_local\_gal} & nominal & is within 10 kpc or 3 Petrosian radii of a galaxy in the 200 Mpc sample?\\
{\tt apparently\_circumnuclear} & nominal & is the source consistent with occurring at the \\
  & &  center of a local universe galaxy?\\
\enddata
\tablenotetext{a}{\url{http://www.usno.navy.mil/USNO/astrometry/optical-IR-prod/icas/icas-usno-b1-format}}
\tablenotetext{b}{From the {\tt sppParams} table of SDSS. See \citet{2004AJ....128..585Y}.}
\tablenotetext{c}{See \url{http://www.sdss.org/dr7/products/spectra/spectroparameters.html}.}
\tablenotetext{d}{See \url{http://cas.sdss.org/astrodr7/en/help/browser/enum.asp?n=SpeczWarning}.}
\label{tab:contextfeatures}
\end{deluxetable}%

\begin{deluxetable}{lcccccc} 
\rotate
\tablecolumns{7} 
\tablewidth{7.5in}
\tablecaption{Oarical Discovery and Classification Statistics} 
\tablehead{ 
\colhead{PTF Type} &  \colhead{Oarical\tablenotemark{a}} & \colhead{Human\tablenotemark{b}} & \colhead{Oarical-Only\tablenotemark{c}} & \colhead{Human\tablenotemark{d}} & \colhead{Oarical\tablenotemark{e}} & \colhead{Human Different\tablenotemark{f}} \\
\colhead{...robotclass} &  \colhead{Discovery} & \colhead{Discovery} & \colhead{Discovery} & \colhead{Rediscovery} & \colhead{Rediscovery} & \colhead{Type}}
\startdata
VarStar & 8322 & 2806 & 5516  & 13 & 2793 & 184 \\
... CV  & 271 & \\
... Periodic & 3081  & \\
Transient & 6246 & 1938 & 4308  & 269 & 1669 & 852 \\
... AGN-cnSN-TDE  & 2295 & \\
... QSO & 1059 & \\
... SN/Nova & 2427 & \\

\enddata
\tablenotetext{a}{Total number of autonomous discoveries and identification of PTF type.}
\tablenotetext{b}{Total number of human-scanned discoveries and identification of  PTF type.}
\tablenotetext{c}{Total number of sources where Oarical was the only discoverer.}
\tablenotetext{d}{Number of sources for which human-scanned discovery occured after autonomous Oarical discovery.}
\tablenotetext{e}{Number of sources for which autonomous Oarical discovery occured after human-scanned discovery.}
\tablenotetext{f}{Number of sources for which human-scanned PTF type differs from Oarical-determined type.}
\label{tab:stats}
\end{deluxetable}%

\end{document}